\documentclass[paper,notoc,nohyper]{JHEP}
\usepackage{epsfig}
\usepackage{amsbsy} 
\def\S{S_{\epsilon}}
\def\T{{\cal T}}

\def\Bfin{{\rm Box^6} (u, t) }
\def\Bubl{{\rm Bub}(s)}

\def\lnx{X}
\def\lny{Y}
\def\Ls{S}

\def\Libx{{\rm Li}_2(x)}

\def\Licx{{\rm Li}_3(x)}
\def\Licy{{\rm Li}_3(y)}
\def\Lidx{{\rm Li}_4(x)}
\def\Lidy{{\rm Li}_4(y)}
\def\Lidz{{\rm Li}_4\Biggl(\frac{x-1}{x}\Biggr)}

\def\tou{\frac{t^2}{u^2}}

\def\CA{C_A}
\def\CF{C_F}

\def\TR{T_R}
\def\NF{N_F}
\def\M{{\cal M}}
\def\A{{\cal A}}

\renewcommand\O[1]{{\cal O}\left(#1\right)}
\def\as{\ensuremath{\alpha_{s}}}
\def\a0{\alpha_0}

\def\beq{\begin{equation}}
\def\eeq{\end{equation}}

\def\beqn{\begin{eqnarray}}
\def\eeqn{\end{eqnarray}}

\def\lq{\left[}
\def\rq{\right]}

\def\({\left(}
\def\){\right)}

\def\ket#1{|{#1}\rangle}
\def\bra#1{\langle{#1}|}
\def\braket#1#2{\langle #1 |#2 \rangle}

\def\MSbar{$\overline{{\rm MS}}$}

\def\fs{\(-\frac{\mu^2}{s}\)^\ep }

\def\fu{\(-\frac{\mu^2}{u}\)^\ep }

\def \ep{\epsilon}

\def\Ls{L_s}
\def\Lu{L_u}
\def\Lx{L_x}
\def\Ly{L_y}
\def\Lx{X}
\def\Ly{Y}
\def\Ls{S}
\def\Lu{U}
\def\Libx{{\rm Li}_2(x)}

\def\Licx{{\rm Li}_3(x)}
\def\Licy{{\rm Li}_3(y)}

\def\Lidx{{\rm Li}_4(x)}
\def\Lidy{{\rm Li}_4(y)}
\def\Lidz{{\rm Li}_4(z)}

\def\Lx{X}
\def\Ly{Y}
\def\Ls{S}
\def\Lip#1#2{{\rm Li}_{#1}(#2)}
\def\tou{\frac{t}{u}}
\def\ttoss{\frac{t^2}{s^2}}
\def\tttosss{\frac{t^3}{s^3}}
\def\sssottt{\frac{s^3}{t^3}}
\def\ttouu{\frac{t^2}{u^2}}
\def\tttouuu{\frac{t^3}{u^3}}
\def\one{}

\def\LU{U}
\def\tpsou{\frac{t^2+s^2}{u^2}}
\def\tpsost{\frac{t^2+s^2}{st}}
\def\tmsou{\frac{t^2-s^2}{u^2}}
\def\tmsost{\frac{t^2-s^2}{st}}
\def\ssott{\frac{s^2}{t^2}}

\def\Ione#1{\mbox{\boldmath I}_{#1}^{(1)}(\ep)}
\def\Iee#1{\mbox{\boldmath I}_{#1}^{(1)}(2\ep)}
\def\Itwo#1{\mbox{\boldmath I}_{#1}^{(2)}(\ep)}

\def\qqg{{{q \bar q g \gamma}}}
\def\qq{{{q \bar q \gamma \gamma}}}
\def\ee{{{e^- e^+ \gamma \gamma}}}

\def\Poltwo#1{{{\cal P}_{\!oles,}^{\, 0 \times 2}}_{#1}}
\def\Fintwo#1{{{\cal F}_{\!inite,}^{\, 0 \times 2}}_{#1}}
\def\Polone#1{{{\cal P}_{\!oles,}^{\, 1 \times 1}}_{#1}}
\def\Finone#1{{{\cal F}_{\!inite,}^{\, 1 \times 1}}_{#1}}

\def\lo{{\rm LO}}
\def\nlo{{\rm NLO}}
\def\nnlo{{\rm NNLO}}
\def\proc{{\cal P}}

\title{\boldmath Two-loop QED and QCD corrections to massless fermion-boson  
scattering\footnote{Work supported in part by the UK Particle Physics and
Astronomy Research Council, by the EU Fourth Framework Programme
`Training and Mobility of Researchers', Network `Quantum Chromodynamics
and the Deep Structure of Elementary Particles',
contract FMRX-CT98-0194 (DG 12 - MIHT) and by the 
National Science Foundation, grant PHY9722101.
}}
\author{
C.~Anastasiou$^{a,b}$,
E.~W.~N.~Glover$^a$,
and M.~E.~Tejeda-Yeomans$^{a,c}$\\
$^a$Department of Physics, 
University of Durham, 
Durham DH1 3LE, 
England\\[1mm]
$^b$ Theory Group, MS81, 
SLAC, 2575 Sand Hill Rd,
Menlo Park, CA 94025, 
U.S.A.\\[1mm]
$^c$C.N.~Yang Institute for Theoretical Physics,
State University of New York,
Stony Brook, NY 11794-3840.
U.S.A.\\[1mm]
E-mail: \email{babis@slac.stanford.edu}, \email{E.W.N.Glover@durham.ac.uk}, 
\email{tejeda@insti.physics.sunysb.edu}}
\abstract{ 
We present the NNLO QCD virtual corrections for 
$q \bar q \to g \gamma,$
$q \bar q \to \gamma \gamma$
and the NNLO QED virtual corrections for 
$e^- e^+\to \gamma \gamma$, 
and all processes related by crossing symmetry. 
We perform an explicit evaluation of the two-loop 
diagrams in  conventional dimensional regularisation, and our 
results are renormalised in the \MSbar\ scheme. The infrared pole structure 
of the amplitudes is in agreement with the prediction of Catani's general 
formalism for the singularities of two-loop amplitudes, while expressions 
for the finite remainder are given for all processes in terms of logarithms 
and polylogarithms  that are real in the physical region.}

\preprint{{DCPT/02/10}, {IPPP/02/05}, {YITP-SB-02-04}, {SLAC-PUB-9130}, {hep-ph/0201274}}

\begin{document}

\section{Introduction}
\label{sec:intro}

Scattering processes involving either initial or final state photons are
common in electron-positron annihilation, electron-proton collisions  and
hadron-hadron collisions.  In the initial state, we may be interested in
direct processes where the photon behaves in a pointlike way, or in
processes where the internal structure of the photon is probed.
Similarly,  prompt final state photons may be produced directly at large
transverse momentum in the hard scattering  or from the fragmentation of
a large transverse momentum jet.\footnote{Non-prompt photons may also be
produced  from the decay of a single large transverse momentum hadron
such as the $\pi^0$.}  

To illustrate the wide range of interesting processes involving photons, 
let us select a few examples relevant to
$e^-e^+$, $e^- p$, and hadron-hadron collider  experiments in the next decade or
so.
In hadron-hadron collisions, attention
is mainly focussed on single direct photon production, 
$$
q \bar q \to \gamma g, \qquad qg \to \gamma q$$
which is sensitive to the
gluonic content of the proton and is an important test of perturbative QCD 
and the pair production of high mass prompt photon
pairs,
$$
q\bar q \to \gamma \gamma, \qquad gg \to \gamma \gamma$$
 which is a background to the discovery of a light Higgs boson via its
decay into photons at the TEVATRON or LHC,
$$
gg \to H \to \gamma \gamma.
$$  
Important processes in electron-positron collisions include
jet photoproduction,
$$
\gamma q \to qg, \qquad \gamma g \to q\bar q,
$$
which is sensitive to the value of the strong coupling constant,
as well as the photoproduction of prompt photons from both the direct
$$
\gamma q \to \gamma q,$$
and resolved processes such as
$$
g q \to \gamma q,
$$
which is sensitive to the gluonic content of the photon.
Finally, the proposed gamma-gamma option of future linear colliders will,
amongst other things, 
measure dijet production,
$$\gamma \gamma \to q\bar q$$
which is once again a background to the Higgs boson detection via its decay into
bottom quarks,
$$
\gamma\gamma \to H \to b\bar b.$$
 
State-of-the-art theoretical predictions for prompt photon  production
and the vast majority of QED and QCD scattering processes,  incorporate
corrections at next-to-leading order (NLO) in perturbation  theory. 
However, these NLO calculations generally exhibit a large sensitivity to
the variation of unphysical renormalisation  and factorization scales  as
a consequence of truncating the perturbative expansion.  The inclusion of
higher order terms is therefore  desirable in order to stabilise the
theoretical predictions and to reduce the inherent theoretical
uncertainties. 

Until recently, a major stumbling block toward a next-to-next-to-leading  order
(NNLO) numerical prediction for scattering processes has been the evaluation of
the two-loop matrix elements. The first calculation of a two-loop four-point
scattering amplitude  was performed by Bern, Dixon, and Kosower  for the
maximal-helicity-violating gluon-gluon scattering~\cite{BDK}.  Subsequently,
generic $2 \to 2$ scattering  matrix elements at two-loops  have now become
tractable for massless particle exchanges in the loops and where all of the
external particles are on-shell. Smirnov~\cite{planarA} and
Tausk~\cite{nonplanarA} have provided analytic  expansions in
$\ep=\frac{4-D}{2}$, where $D$ is the dimension, in terms of Nielsen
polylogarithms for the two most  challenging integrals emerging; the
double-box~\cite{planarA} and the cross-box~\cite{nonplanarA}.   At the same
time, algorithms based on integration by parts~\cite{IBP} and Lorentz
invariance~\cite{diffeq} recursion relations were also developed for the tensor
reduction to master integrals of all relevant two-loop
topologies~\cite{planarB, nonplanarB, AGO3,onshell5,onshell6}.

Already, this technology has been applied to a wide range of physically
interesting processes.  The interference of tree and two-loop graphs
(together with the simpler self interference of one-loop diagrams) for 
various processes  have now been computed, including Bhabha scattering
($e^+e^- \to e^+e^-$) in the massless electron limit~\cite{BDG} and all
the QCD $2 \to 2$ parton-parton  scattering processes ($gg \to gg$, $gg
\to q\bar q$ and $q \bar q \to q\bar q$)~\cite{qqQQ, qqqq, 1loopsquare, 
qqgg, gggg, 1loopgggg}. Two-loop helicity amplitudes have also been
derived for gluon fusion into photons ($g g \to \gamma \gamma$)
~\cite{ggpp}, light-by-light scattering ($\gamma \gamma \to \gamma
\gamma$)~\cite{pppp} and gluon-gluon scattering ($gg \to
gg$)~\cite{gggg_BDD}.

The case where the internal propagators are massless but one external leg
is off-shell has also been intensively studied, leading to the evaluation
of all associated planar and non-planar master
integrals~\cite{mi,smirnov_offshell} in terms of two-dimensional harmonic
polylogarithms~\cite{hpl}.  These integrals arise in the decay of an
off-shell photon to three partons ($\gamma^* \to q\bar q g$) relevant for
three jet production in electron-positron annihilation and the NNLO
matrix elements have been evaluated in Ref.~\cite{zqqg}. Even more
recently, Mellin Barnes integral techniques have been applied to the
planar double box diagram with four massive and three massless lines, 
all four legs on the mass shell~\cite{smirnov_mass}.

In this paper, we present the NNLO virtual corrections for the 
processes  \begin{eqnarray} && q   \;+\;  \bar q \;\to \;    g    \;+\;
\gamma, \nonumber \\ && q   \;+\;  \bar q \;\to \;  \gamma \;+\; \gamma,
\nonumber \\ && e^- \;+\;   e^+   \;\to \;  \gamma \;+\; \gamma,
\nonumber  \label{eq:processes} \end{eqnarray}
and the ones related by crossing symmetry or time-reversal. For clarity
and calculational convenience, we decompose the NNLO  virtual corrections
into the self-interference of the one-loop  amplitude and the
interference of the tree and two-loop amplitude.    Our results are valid
in the  limit where the masses of the quarks and electrons can be
ignored.  The Feynman diagrams for the processes in consideration are
only a subset of  those for quark-gluon scattering ($q \bar q \to gg$),
which we presented in~\cite{qqgg}. However, due to the complicated colour
structure of the  quark-gluon two-loop amplitude and the different
flavour content of the  processes with photons we choose to present them
independently.  As in~\cite{qqgg}, we work in dimensional regularisation
treating all external particles in $D$ dimensions, and renormalise with
the \MSbar\  scheme.  

Our paper is organised as follows. In Section~\ref{sec:notation} we
introduce our notation and define the perturbative expansion of the
matrix elements summed over colours and spins.  In
Section~\ref{sec:infra} we study the singular behavior of   the NNLO
contributions, and verify that it agrees with the general formalism 
developed by Catani for the infrared structure of two-loop 
amplitudes~\cite{catani}.  For the simpler case of one-loop  amplitudes,
their poles in $\ep$ can be  expressed in terms of the tree  amplitude 
and a colour-charge operator $\Ione{}$, constructed  in a  universal
manner~\cite{seymour}.  In the same fashion, the divergences of the
two-loop amplitude can be  written as a sum of two terms: the action of
the $\Ione{}$ operator on the  one-loop amplitude and the action of a new
operator $\Itwo{}$ on the tree  amplitude. The $\Itwo{}$ operator
includes a renormalisation scheme  dependent term $H^{(2)}$ multiplied by
a $1/\ep$ pole. Although  in~\cite{catani} it is  anticipated that
$H^{(2)}$  can be constructed for  any given process in a universal
manner, such a prescription is not yet  available.  We give explicit
expressions for $\Ione{}$ and $\Itwo{}$ relevant for each of the
processes in Eq.~(\ref{eq:processes}) valid in the \MSbar\ scheme. In
Section~\ref{sec:two} we present the finite remainders  of the
interference of the tree and the two-loop amplitude after  subtraction of
the singular poles of Section~\ref{sec:infra} from the explicit  result
of the two-loop Feynman diagrams. We organize the finite part  according
to  the colour and flavour content of the two-loop amplitude.  Similarly,
in  Section~\ref{sec:one} we give the finite contributions of  the 
self-interference of the one-loop amplitude. The finite remainders are 
expressed in terms of logarithms and polylogarithms which are real in
the  physical domain. Finally, we conclude in Section~\ref{sec:conc}.

\newpage
\section{Notation}
\label{sec:notation}

We consider the processes 
\begin{eqnarray}
&& 
\label{eq:qqg}
q(p_1) + \bar q(p_2) +g(p_3) +\gamma(p_4) \to 0, \\
&& 
\label{eq:qq}
q(p_1) + \bar q(p_2) +\gamma(p_3) +\gamma(p_4) \to 0, \\
&& 
\label{eq:ee}
e^-(p_1) + e^+(p_2) +\gamma(p_3) +\gamma(p_4) \to 0,
\end{eqnarray}
where all particles are incoming with light-like momenta 
satisfying 
\begin{equation}
p_1^\mu+p_2^\mu+p_3^\mu+p_4^\mu=0, \quad p_i^2=0. 
\end{equation} 

The amplitudes exhibit infrared and ultraviolet divergences so we work in 
conventional dimensional regularisation and treat all external particle  
states in $D$ dimensions.  The ultraviolet divergences are renormalised 
with the \MSbar\ scheme where the bare strong coupling $\alpha_s^0$ is related to 
the running coupling  $\as \equiv \alpha_s(\mu^2)$  
at renormalisation scale $\mu$ via
\begin{equation}
\label{eq:alpha}
\alpha_s^0 \,  \S = \as \,  \lq 1 - \frac{\beta_0}{\ep}  
\, \left(\frac{\as}{2\pi}\right) + \( \frac{\beta_0^2}{\ep^2} - \frac{\beta_1}{2\ep} \)  \, \left(\frac{\as}{2\pi}\right)^2
+\O{\as^3} \rq.
\end{equation}
In this expression
\begin{equation}
\S = (4 \pi)^\ep e^{-\ep \gamma},  \quad\quad \gamma=0.5772\ldots=
{\rm Euler\ constant}
\end{equation}
is the typical phase-space volume factor in $D=4-2\ep$ dimensions and
$\beta_0, \beta_1$ are the first two coefficients of the QCD beta function for $\NF$ (massless) quark flavours
\begin{equation}
\label{betas}
\beta_0 = \frac{11 \CA - 4 T_R \NF}{6} \;\;, \;\; \;\;\;\;
\beta_1 = \frac{17 \CA^2 - 10 \CA T_R \NF - 6 \CF T_R \NF}{6} \;\;.
\end{equation}
For an $SU(N)$ gauge theory
\begin{equation}
\CF = \frac{N^2-1}{2N}, \qquad \CA = N, \qquad T_R = \frac{1}{2}.
\end{equation}

In a similar way,  the bare QED coupling $\alpha^0$ is related to 
the running coupling  $\alpha \equiv \alpha({\mu^\prime}^2)$  
at renormalisation scale $\mu^\prime$ via
\begin{equation}
\label{eq:alpha_QED}
\alpha^0 \,  \S = \alpha \,  \lq 1 - \frac{\beta_0^\prime}{\ep}  
\, \left(\frac{\alpha}{2\pi}\right) + \( \frac{{\beta_0^\prime}^2}{\ep^2} - \frac{\beta_1^\prime}{2\ep} \)  \, \left(\frac{\alpha}{2\pi}\right)^2
+\O{\alpha^3} \rq.
\end{equation}
where
$\beta_0^\prime, \beta_1^\prime$ are now the first two coefficients of the QED beta function,
\begin{equation}
\label{betasQED}
\beta_0^\prime = \frac{ - 2 \NF^\prime}{3} \;\;, \;\; \;\;\;\;
\beta_1^\prime = - \NF^\prime \;\;,
\end{equation}
and  
\begin{equation}
\NF^\prime=\sum_f Q_f^2,
\end{equation}
with the sum running over the active (massless) fermion flavours.

The renormalised amplitudes may be expanded as 
\begin{eqnarray}
\ket{\M_\qqg} &=& 4 \pi \sqrt{ \alpha \as} \lq \ket{\M_\qqg^{(0)}}
+ \left( \frac{\as}{2\pi}\right) \ket{\M_\qqg^{(1)}}
+ \left( \frac{\as}{2\pi}\right)^2 \ket{\M_\qqg^{(2)}}
+ \O{\as^3 \alpha^0}
\rq, \nonumber \\
&& \\
\ket{\M_\qq} &=& 4 \pi \alpha  \lq \ket{\M_\qq^{(0)}}
+ \left( \frac{\as}{2\pi}\right) \ket{\M_\qq^{(1)}}
+ \left( \frac{\as}{2\pi}\right)^2 \ket{\M_\qq^{(2)}}
+ \O{\as^3 \alpha^0}
\rq, \nonumber \\
&& \\
\ket{\M_\ee} &=& 4 \pi \alpha  \lq \ket{\M_\ee^{(0)}}
+ \left( \frac{\alpha}{2\pi}\right) \ket{\M_\ee^{(1)}}
+ \left( \frac{\alpha}{2\pi}\right)^2 \ket{\M_\ee^{(2)}}
+ \O{\alpha^3}
\rq, \nonumber \\
\end{eqnarray}
where $\ket{\M_\proc^{(i)}}$ represents the colour-space vector describing 
the renormalised $i-$loop amplitude for the $\proc=\qqg, \, \qq, \,\ee$ 
processes  of Eqs.~(\ref{eq:qqg})-~(\ref{eq:ee}).
The dependence on the renormalisation scale $\mu$ and renormalisation scheme is implicit.

We denote the squared amplitudes summed over spins and colours by 
\begin{eqnarray}
\label{eq:processA}
\braket{\M_\qqg}{\M_\qqg} &=& \sum \left|\M\left( 
q + \bar q \to \gamma + g
\right) \right|^2 = \A_\qqg\left( s, t, u \right), \\
\braket{\M_\qq}{\M_\qq} &=& \sum \left|\M\left( 
q + \bar q \to \gamma + \gamma
\right) \right|^2 = \A_\qq\left( s, t, u \right), \\
\braket{\M_\ee}{\M_\ee} &=& \sum \left|\M\left( 
e^- + e^+ \to \gamma + \gamma
\right) \right|^2 = \A_\ee\left( s, t, u \right), 
\end{eqnarray} 
where the Mandelstam variables are
\begin{equation}
s=(p_1+p_2)^2, \quad t=(p_2+p_3)^2, \quad u=(p_1+p_3)^2, \quad 
s+t+u=0. 
\end{equation}

The squared matrix elements for the crossed processes  
are obtained by permuting the Mandelstam variables and introducing 
a minus sign for each fermion exchange between initial and final states,  
\begin{eqnarray}
\sum |{\cal M}({g + \gamma \to  q + \bar q })|^2  
&=& \A_\qqg(s,t,u),\\
\sum |{\cal M}({q + \gamma \to  q + g })|^2  
&=& - \A_\qqg(u,t,s),\\
\sum |{\cal M}({q + g      \to  \gamma + q  })|^2  
&=& - \A_\qqg(t,u,s),\\
\sum |{\cal M}({g + \bar q \to  \gamma + \bar q })|^2  
&=& - \A_\qqg(u,t,s), \\
\sum |{\cal M}({\gamma + \bar q \to  \bar q+ g  })|^2  
&=& - \A_\qqg(t,u,s).
\end{eqnarray}
Similarly, 
\begin{eqnarray}
\sum |{\cal M}({\gamma + \gamma \to  q + \bar q })|^2  
&=& \A_\qq(s,t,u),\\
\sum |{\cal M}({q + \gamma \to  q + \gamma })|^2  
&=& - \A_\qq(u,t,s),\\
\sum |{\cal M}({\gamma + \bar q \to  \gamma + \bar q })|^2  
&=& - \A_\qq(u,t,s), 
\end{eqnarray}
and 
\begin{eqnarray}
\sum |{\cal M}({\gamma + \gamma \to  e^- + e^+ })|^2  
&=& \A_\ee(s,t,u),\\
\sum |{\cal M}({e^- + \gamma \to  e^- + \gamma })|^2  
&=& - \A_\ee(u,t,s),\\
\label{eq:processB}
\sum |{\cal M}({\gamma + e^+ \to  \gamma + e^+ })|^2  
&=& - \A_\ee(u,t,s). 
\end{eqnarray}

The functions $\A_\proc (s, t, u)$ are symmetric under the exchange of $t$ 
and $u$ and can be expanded perturbatively to yield, 
\begin{eqnarray}
\A_\qqg \left(s,t,u\right)=16\pi^2\alpha \as 
&\Bigg[&
\A_\qqg^\lo\left(s,t,u\right) 
+ \left(\frac{\as}{2\pi}\right) \, \A_\qqg^\nlo \left(s,t,u\right)
 \nonumber \\
&&+ \left(\frac{\as}{2\pi}\right)^2 \A_\qqg^\nnlo \left(s,t,u\right)
+ \O{\as^3 \alpha^0}
\Bigg],
\end{eqnarray}   
\begin{eqnarray}
\A_\qq \left(s,t,u\right)=16\pi^2\alpha^2 
&\Bigg[&
\A_\qq^\lo \left(s,t,u\right) 
+ \left(\frac{\as}{2\pi}\right) \,  \A_\qq^\nlo \left(s,t,u\right)
 \nonumber \\
&&+ \left(\frac{\as}{2\pi}\right)^2 \A_\qq^\nnlo \left(s,t,u\right)
+ \O{\as^3 \alpha^0}
\Bigg],
\end{eqnarray}   
and 
\begin{eqnarray}
\A_\ee \left(s,t,u\right)=16\pi^2\alpha^2 
&\Bigg[&
\A_\ee^\lo \left(s,t,u\right) 
+ \left(\frac{\alpha}{2\pi}\right) \, \A_\ee^\nlo \left(s,t,u\right)
 \nonumber \\
&&+ \left(\frac{\alpha}{2\pi}\right)^2 \A_\ee^\nnlo \left(s,t,u\right)
+ \O{\alpha^3}
\Bigg].
\end{eqnarray}   

At leading-order (LO) the self-interference of the tree amplitudes is
\begin{equation}
\A_\proc ^\lo \left(s,t,u\right) = 
\braket{\M_\proc^{\left(0\right)}}{\M_\proc^{\left(0\right)}},
\end{equation}
so that for each process we have
\begin{eqnarray}
\label{eq:tree_qqg}
\A_\qqg^\lo (s, t, u) &=& N \CF  \T(s, t, u),\\
\label{eq:tree_qq}
\A_\qq^\lo (s, t, u)  &=& N \T(s, t, u),\\
\label{eq:tree_ee}
\A_\ee^\lo (s, t, u)  &=& \T(s, t, u),
\end{eqnarray}
where we have defined the tree-type structure
\begin{equation}
\T(s, t, u) = 8\left( 1- \ep \right) 
\left( \frac{u}{t} + \frac{t}{u} -\ep \frac{s^2}{tu} \right).
\end{equation}

The next-to-leading order (NLO) term consists of the interference of the tree
and the one-loop amplitudes
\begin{equation}
\A_\proc ^\nlo \left(s,t,u\right) =  
\braket{\M_\proc^{\left(0\right)}}{\M_\proc^{\left(1\right)}} 
+\braket{\M_\proc^{\left(1\right)}}{\M_\proc^{\left(0\right)}},  
\end{equation}
and their expansion in $\ep$ up to $\O{\ep}$ was given in 
Ref.~\cite{oneloopME}. 
As will be discussed in Section~\ref{sec:infra}, the singular structure of the 
two-loop amplitude is expressed in terms of the tree amplitude multiplied 
by a singular operator of $\O{1/\ep^4}$ and the one-loop amplitude multiplied by another
operator diverging as $1/\ep^2$. Therefore, one needs to know the expansion 
of the one-loop amplitude up to and including the $\O{\ep^2}$ term. In 
Section~\ref{sec:infra} we give expressions for 
$ \braket{\M_\proc^{\left(0\right)}}{\M_\proc^{\left(1\right)}} $ valid in 
all kinematic regions and to all orders in $\ep$, in terms of two one-loop 
master integrals. The $\ep$-expansions of these master integrals to the 
appropriate order are given in the Appendix~\ref{app:master_int}.  

At next-to-next-to-leading-order (NNLO)  contributions from 
the self-interference of the one-loop amplitude and the interference of 
the tree and the two-loop amplitude must be taken into account, so that
\begin{equation}
\A_\proc ^\nnlo (s, t, u) =\A_\proc^{\nnlo (1 \times 1)}(s, t, u)
+ \A_\proc^{\nnlo (0 \times 2)}(s, t, u),
\end{equation}
with 
\begin{equation}
\A_\proc^{\nnlo (1 \times 1)}(s, t, u)=\braket{\M_\proc^{(1)}}{\M_\proc^{(1)}},
\end{equation}
and 
\begin{equation}
\A_\proc^{\nnlo (0 \times 2)}(s, t, u)=\braket{\M_\proc^{(0)}}{\M_\proc^{(2)}}
  +\braket{\M_\proc^{(2)}}{\M_\proc^{(0)}}.
\end{equation}
In the following sections, we present expressions for the infrared singular 
and finite contributions to $\A_\proc^\nnlo$. In Section~\ref{sec:infra} 
we write down the singular parts according to the formalism of 
Catani~\cite{catani}. 
The pure two-loop finite contributions 
of $\A_\proc^{\nnlo \, (2 \times 0)}(s,t,u) $ are described in 
Sec.~\ref{sec:two} and the finite remainders of the self-interference of 
the one-loop amplitude $\A_\proc^{\nnlo\, (1 \times 1)}(s,t,u)$ are described 
in Sec.~\ref{sec:one}.

As in Refs.~\cite{qqQQ,qqqq, 1loopsquare,qqgg, gggg,1loopgggg}, 
we use {\tt QGRAF}~\cite{QGRAF} to  produce the
tree, one and two-loop Feynman diagrams to construct $\ket{\M_\proc^{(i)}}$. 
We then project by 
$\bra{\M_\proc^{(0)}}$ or $\bra{\M_\proc^{(1)}}$
and perform the summation over colours and
spins. It should be noted that when summing over the gluon polarisation, 
we ensure
that the polarisation states are transversal (i.e. physical) by using 
an axial gauge
\begin{equation}
\sum_{{\rm spins}} \ep_3^{\mu}\ep_3^{\nu\ast} = 
-  g^{\mu \nu} + \frac{n^{\mu}p_3^{\nu} 
+ n^{\nu}p_3^{\mu}}{n \cdot p_3}, 
\end{equation}
where $n$ is an arbitrary light-like 4-vector. 
For simplicity, we choose $n^{\mu} = p_4^{\mu}$.
Finally, the  trace over the Dirac matrices is carried  out in $D$
dimensions using conventional dimensional regularisation. It is then
straightforward to identify the scalar and tensor integrals present  and
replace them with combinations of the basis set  of master integrals using
the  tensor reduction of two-loop integrals described in
\cite{planarB,nonplanarB,AGO3}, based on integration-by-parts~\cite{IBP} and 
Lorentz invariance~\cite{diffeq} identities.   The final result is  a
combination of master integrals in $D=4-2\epsilon$ for which the 
expansions around $\ep = 0$ are given
in~\cite{planarA,nonplanarA,planarB,nonplanarB,AGO3,onshell5,onshell6,AGO2,xtri}.  

\section{Infrared Pole Structure}
\label{sec:infra}
We further decompose the one-loop self-interference and the 
two-loop contributions as a sum of singular and finite terms, 
\begin{equation}
\A_\proc^{\nnlo \, (1\times 1)}(s,t,u)
 = \Polone{\proc}(s,t,u)+ \Finone{\proc}(s,t,u)
\end{equation} 
and
\begin{equation}
\A_\proc^{\nnlo \, (2\times 0)}(s,t,u)
 = \Poltwo{\proc}(s,t,u)+ \Fintwo{\proc}(s,t,u),
\end{equation} 
for each of the processes $\proc=\qqg, \, \qq, \, \ee.$
$\Polone{\proc}$ and $\Poltwo{\proc}$ contain infrared singularities that 
will be analytically canceled by the infrared singularities occurring in radiative processes of the same order 
(ultraviolet divergences having already being removed by 
renormalisation). $\Finone{\proc}$ and $\Fintwo{\proc}$ are the remainders 
which are finite as $\ep \to 0$.

The poles of the one-loop amplitude 
self-interference can be written in terms of a universal operator 
$\Ione{}$ acting on the colour-space of the amplitude. Due to the 
simple colour structure of the processes we study, the action of 
$\Ione{}$ factorises yielding, 
\begin{equation}
\Polone{\proc}(s,t,u)=-\left|\Ione{\proc} \right|^2 
\braket{\M_\proc^{(0)}}{\M_\proc^{(0)}}
+ 2 {\rm Re} \left\{ \Ione{\proc}^\dagger 
\braket{\M_\proc^{(0)}}{\M_\proc^{(1)}}\right\} 
\end{equation} 
where 
\begin{eqnarray}
\Ione{\qqg} = \frac{e^{\ep \gamma}}{2 \Gamma(1-\ep)} &&
\Bigg[ 
-N \left(\frac{1}{\ep^2} + \frac{3}{4\ep} + \frac{\beta_0}{2 N \ep} \right) 
\left\{
\left(-\frac{\mu^2}{t}\right)^\ep +
\left(-\frac{\mu^2}{u}\right)^\ep \right\} 
\nonumber \\
&& + \frac{1}{N}\left(\frac{1}{\ep^2} + \frac{3}{2\ep}\right) 
\left(-\frac{\mu^2}{s}\right)^\ep \Bigg],
\label{eq:Ioneqqg}
\end{eqnarray}
\begin{eqnarray}
\Ione{\qq} = -\CF \frac{e^{\ep \gamma}}{\Gamma(1-\ep)} 
\left(\frac{1}{\ep^2} + \frac{3}{2\ep} \right) 
\left(-\frac{\mu^2}{s}\right)^\ep
\label{eq:Ioneqq}
\end{eqnarray}
and 
\begin{eqnarray}
\Ione{\ee} = -\frac{e^{\ep \gamma}}{\Gamma(1-\ep)} 
\left(\frac{1}{\ep^2} + \frac{3}{2\ep} 
 + \frac{\beta_0^\prime}{ \ep} \right) 
\left(-\frac{\mu^2}{s}\right)^\ep.
\label{eq:Ioneee}
\end{eqnarray}
The generic form of the $\Ione{}$ operator was found by Catani and 
Seymour~\cite{seymour} and it was derived for the general one-loop QCD 
amplitude by integrating the real radiation graphs of the same order  in
perturbation series in the one-particle unresolved limit. For the 
processes $\qqg$ and $\qq$ we consider only NNLO QCD corrections,
therefore  in Eq.~(\ref{eq:Ioneqq}) there are no contributions from soft or
collinear photon emission at the same order in $\as$.  Note that on
expanding the $\Ione{}$ operator, imaginary parts are generated, the sign
of which are fixed by the small imaginary part $+i0$ assigned to each
Mandelstam variable. Terms that involve the hermitian conjugate of this
operator are to be modified accordingly.

The renormalised interference of the tree and the one-loop amplitudes can be expressed 
in terms of the one-loop bubble graph (Bub) and the one-loop box integral in
$D=6-2\ep$ dimensions (Box$^6$), as follows
\begin{eqnarray}
\braket{\M_\qqg^{(0)}}{\M_\qqg^{(1)}} &=&  N \CF \, \Bigg[ 
\CF f_1\left(s, t, u\right) - \frac{\CA}{2} \left\{ 
f_1\left(s, t, u \right)  +
f_2\left(s, t, u \right)  \right\}
\nonumber \\
&&
+
\left( t \leftrightarrow u \right)
\Bigg] 
-\frac{\beta_0}{2 \ep} \braket{\M_\qqg^{(0)}}{\M_\qqg^{(0)}},  
\end{eqnarray}

\begin{equation}
\braket{\M_\qq^{(0)}}{\M_\qq^{(1)}} 
=  N \CF f_1(s, t, u) + \left( t \leftrightarrow u\right)
\end{equation}
and 
\begin{equation}
\braket{\M_\ee^{(0)}}{\M_\ee^{(1)}} 
=   f_1(s, t, u) + f_1(s, u, t) 
-\frac{\beta_0^\prime}{\ep} \braket{\M_\ee^{(0)}}{\M_\ee^{(0)}} ,
\end{equation}
where we have used
\begin{eqnarray}
f_1(s, t, u) &=& -\frac{8\left(1-2\ep\right)
\lq
s^2+t^2-\ep \left(s^2+t^2+u^2\right) 
+\ep^2 \left(t^2+s^2-st\right) -\ep^3 st  
\rq
}{t} {\rm Box^6}(s, t) \nonumber \\
&& +\frac{4 \left(1-2\ep \right) \lq 
t^2+u^2+\ep(\ep-2)s^2
\rq
}{\ep ut} {\rm Bub}(s) \nonumber \\
&& -\frac{4(1-\ep) \lq 3s+t+\ep (5t+2u)-3\ep^2 s\rq }{t}
{\rm Bub}(t)
\end{eqnarray}
and
\begin{eqnarray}
f_2(s, t, u) &=& \frac{4 s \left(1-2\ep\right) 
\left[ 
t^2+u^2+ \ep \left( ut-2t^2-2u^2\right) +\ep^2 \left( s^2+tu \right)  
\right] }{ut} {\rm Box^6}\left(t,u\right) \nonumber \\
&& -\frac{4 \left(1-\ep\right)^2 
\left[2\left(t^2+u^2 \right) -\ep \left(t-s\right)^2-\ep^2 u s\right] }{\ep tu}
{\rm Bub}(t).   
\end{eqnarray}
These expressions are valid in all kinematic regions and for all orders in 
$\epsilon$.  
However, to evaluate them in a particular region, the one-loop master 
integrals ${\rm Bub}$ and ${\rm Box}^6$,  must be expanded as a series
in $\epsilon$ (see Appendix~\ref{app:master_int}). 

The infrared poles of the interference of the tree and the two-loop
amplitudes follow a generic formula developed by Catani~\cite{catani},
such that 
\begin{equation}
\Poltwo{\proc}(s,t,u)=2 {\rm Re} \left\{ 
\Ione{\proc} \braket{M_\proc^{(0)}}{M_\proc^{(1)}} 
+ \Itwo{\proc} \braket{M_\proc^{(0)}}{M_\proc^{(0)}} \right\} 
\end{equation} 
where once again the  
simple colour structure of the processes under consideration allows
the action of 
$\Ione{}$ and $\Itwo{}$ to be factrorised. 
For the QCD processes $\proc=\qqg,\, \qq$ we have  
\begin{eqnarray}
\Itwo{\proc} &=& -\frac{1}{2} \Ione{\proc} \left(\Ione{\proc}
+\frac{2\beta_0}{\ep} \right)
+\frac{e^{-\ep \gamma} \Gamma(1-\ep)} {\Gamma(1-2\ep)}
\left(\frac{\beta_0}{\ep}+ K \right) \Iee{\proc} \nonumber \\
&& +\frac{e^{\ep \gamma}}{4 \ep \Gamma(1-\ep)} H_\proc^{(2)}
\end{eqnarray}
with 
\begin{equation}
K=\left(\frac{67}{18} -\frac{\pi^2}{6}\right) \CA -\frac{10}{9} T_R \NF.
\end{equation}
and for the QED process $\proc=\ee$ we have 
\begin{eqnarray}
\Itwo{\ee} &=& -\frac{1}{2} \Ione{\ee} \left(\Ione{\ee}+\frac{2\beta_0^\prime}
{\ep} \right)
+\frac{e^{-\ep \gamma} \Gamma(1-\ep)} {\Gamma(1-2\ep)}
\left(\frac{\beta_0^\prime}{\ep}+ K^\prime \right) \Iee{\ee} \nonumber \\
&& +\frac{e^{\ep \gamma}}{4 \ep \Gamma(1-\ep)} H_\ee^{(2)}
\end{eqnarray}
with 
\begin{equation}
K^\prime=-\frac{10}{9} \NF^\prime.
\end{equation}

The process and renormalisation scheme dependent $H^{(2)}$ constants are
related to the colour space operator $\mbox{\boldmath H}^{(2)}$.   The full
structure of $\mbox{\boldmath H}^{(2)}$ is not known at present, although it
certainly contains non-trivial colour structures ~\cite{gggg_BDD} that
have been investigated in the case of gluon-gluon scattering. However,
because of the projection by the tree-level amplitude and the summation 
over colours we are only sensitive to the trivial colour part of
$\mbox{\boldmath H}^{(2)}$, and we find that the constants $H^{(2)}$ can be
written as a sum of factors associated with each of the  external
legs.    We derive them directly from the $1/\ep$ term of the explicit
expansion in $\ep$ of the two-loop amplitudes.  In conventional
dimensional regularisation and the \MSbar\ scheme, we find that for the
processes under consideration
\begin{eqnarray}
\label{eq:H2qqg}
H_\qqg^{(2)} &=& H_{q}^{(2)}+H_{\bar q}^{(2)} + H_{g}^{(2)}, \\
\label{eq:H2qq}
H_\qq^{(2)} &=& H_{q}^{(2)}+H_{\bar q}^{(2)},\\
H_\ee^{(2)} &=& H_{e-}^{(2)}+H_{e^+}^{(2)}+2 H_{\gamma}^{(2)},
\label{eq:H2ee}
\end{eqnarray}
where for an external quark or an antiquark we find 
\begin{eqnarray}
H_{q}^{(2)} =H_{\bar q}^{(2)} &=&\left(\frac{\pi^2}{2}-6 ~\zeta_3 
-\frac{3}{8}\right) \CF^2
+\left(\frac{13}{2}\zeta_3 +\frac{245}{216}-\frac{23}{48} \pi^2 \right) \CA \CF
\nonumber \\
&& + \left(-\frac{25}{54}+\frac{\pi^2}{12} \right) \TR \NF \CF
\end{eqnarray}
and for each external gluon
\begin{eqnarray}
H_{g}^{(2)} &=& 
\frac{20}{27} \TR^2 \NF^2
+ \TR \CF \NF
-\left(\frac{ \pi^2}{36}+\frac{58}{27} \right)\TR \NF\CA
\nonumber \\ &&
+\left(\frac{\zeta_3}{2}+\frac{5}{12}
+\frac{11}{144}\pi^2 \right) \CA^2 .
\end{eqnarray}
These are the same factors as found in \cite{catani,qqQQ,qqqq,qqgg,gggg,zqqg}.
Similarly, for  each external electron or positron we find 
\begin{eqnarray}
H_{e^-}^{(2)}=H_{e^+}&=&\left(\frac{\pi^2}{2}-6 \zeta_3 -\frac{3}{8}\right) 
+ \left(-\frac{25}{54}+\frac{\pi^2}{12} \right) \NF^\prime
\end{eqnarray}
while for the external photon
\begin{equation}
H_{\gamma}^{(2)} = \frac{20}{27}{\NF^\prime}^2 + \NF^\prime. 
\end{equation}
It should be noted that $H_{e^-}^{(2)}$, $H_{e^+}^{(2)}$ and $H_\gamma^{(2)}$ 
can be obtained 
from the expressions for $H_{q}^{(2)}$,$H_{\bar q}^{(2)}$  and $H_g^{(2)}$ 
accordingly, by taking the limit $\TR \to 1$, $\CF \to 1$ and $\CA \to 0$. 
It should also be emphasized that  contributions of $\O{\ep}$ to 
${H^{(2)}}$ are undetermined at present.    

\section{Finite two-loop contributions}
\label{sec:two}

In this section, we give explicit expressions for the finite remainder of 
the two-loop contributions $\Fintwo{\proc}$ defined as, 
\begin{equation}
 \Fintwo{\proc}(s,t,u) = \A_\proc^{\nnlo \, (2\times 0)}(s,t,u) 
     - \Poltwo{\proc}(s,t,u).
\end{equation} 
Note that the $\Poltwo{\proc}(s,t,u)$ is expanded through to $\O{1}$ and therefore contains finite as well as singular
contributions.
Using the standard polylogarithm identities~\cite{kolbig}, we express 
our results in terms of a basis set of logarithms and polylogarithms\footnote{As usual, the polylogarithms ${\rm Li}_n(w)$ are defined by
\begin{eqnarray}
 {\rm Li}_n(w) &=& \int_0^w \frac{dt}{t} {\rm Li}_{n-1}(t) \qquad {\rm ~for~}
 n=2,3,4\nonumber \\
 {\rm Li}_2(w) &=& -\int_0^w \frac{dt}{t} \log(1-t).
\label{eq:lidef}
\end{eqnarray} }
with arguments $x$, $1-x$ and $(x-1)/x$, where
\begin{equation}
\label{eq:xydef}
x = -\frac{t}{s}, \qquad y = -\frac{u}{s} = 1-x, 
\qquad z=-\frac{u}{t} = \frac{x-1}{x}.
\end{equation}
In the physical region $s>0$ and $t,u<0$, our basis set of functions are all
real. 
For convenience, we also introduce the following logarithms
\begin{equation}
\label{eq:xydef1}
\lnx = \log\left(\frac{-t}{s}\right),
\qquad \lny = \log\left(\frac{-u}{s}\right),
\qquad \Ls = \log\left(\frac{s}{\mu^2}\right),
\qquad \Lu = \log\left(\frac{-u}{\mu^2}\right),
\end{equation}
where $\mu$ is the renormalisation scale.

From Eqs.~(\ref{eq:processA})-(\ref{eq:processB}) we see 
that we need to evaluate 
 the function $\Fintwo{\proc}$ with arguments $(s,t,u)$, $(u,t,s)$
and $(t,u,s)$. Since $t$ and $u$ are both negative they can be interchanged 
leaving the polylogarithms and logarithms of our basis well defined and
 real. Therefore the crossing symmetry 
$\left( t \leftrightarrow u \right)$
can be performed trivially.
It is  a more involved operation obtain expressions valid
under the exchange of $s$ with either 
$t$ or $u$,
since the logarithms and polylogarithms may acquire imaginary parts and 
appropriate analytic continuations need to be considered. 
We therefore give explicit expressions for the ``$s$-channel'' function 
$\Fintwo{\proc}(s, t, u)$ and 
the ``$u$-channel'' function $\Fintwo{\proc}(u, t, s)$
that are directly valid in the physical region $s > 0$ and $u, t < 0$.
The ``t-channel'', 
$\Fintwo{\proc}(t, u, s)$ can be obtained from the $u-$channel expression  
by exchanging $t$ and $u$. Note that here we define a ``channel'' according to 
the first argument of the function $\Fintwo{\proc}$. 

\begin{figure}[t]
\begin{center}
(a)
~\epsfig{file=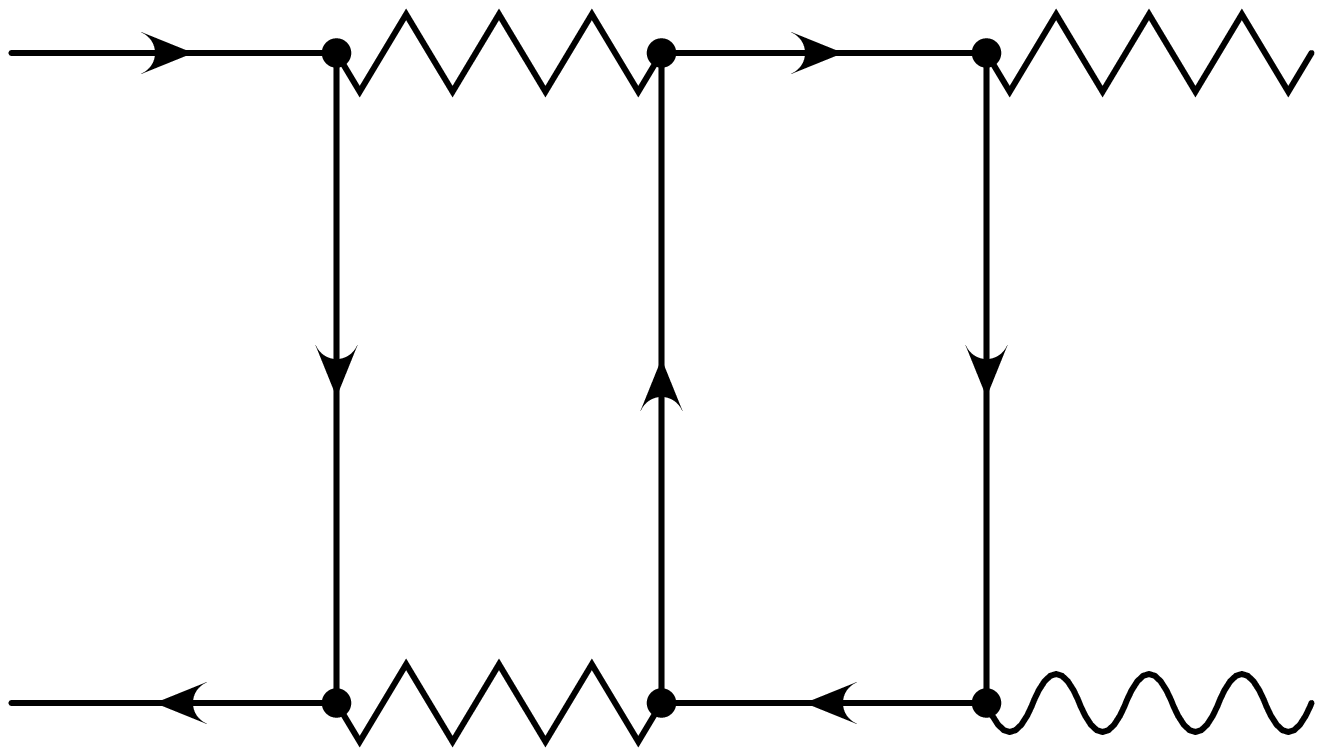,width=6cm}
(b)
~\epsfig{file=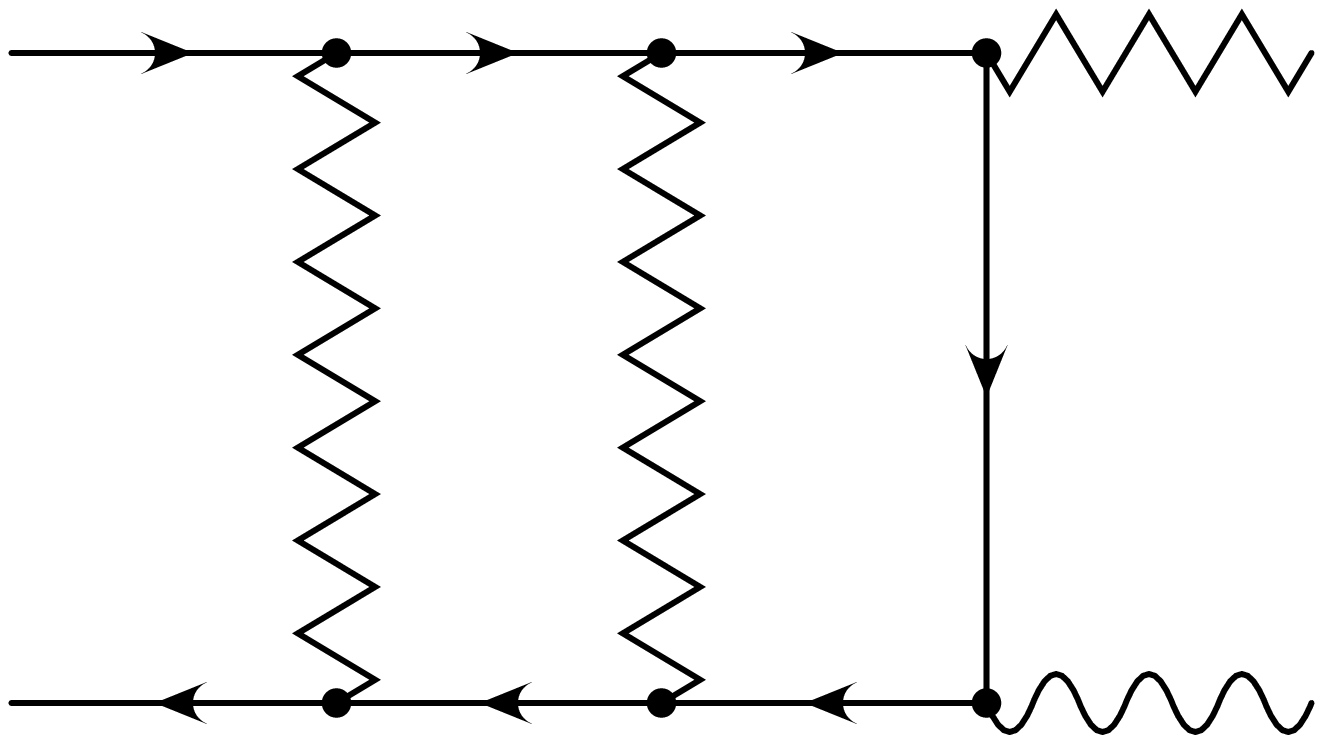,width=6cm}
\end{center}
\caption{Feynman diagrams representative of (a) ``light-by-light"  scattering
and
(b) ``abelian gluon" exchange.   These graphs contribute to $A_c$ and $B_c$
respectively. Depending on the process, the zigzag lines represent either photons
or gluons.}
\label{fig:rep}
\end{figure}

We organize our results according to the colour and flavour structure of the 
two-loop amplitude, so that for the three different processes, we have
\begin{eqnarray}
\Fintwo{\qqg;\; c} &=& 2 N\CF 
\Biggl[
\left( \sum_q Q_q\right) 
\left(
2 \TR \CF - {3\TR \CA\over 4}  
\right) A_c\nonumber \\
&&
+ \CF^2 B_c   
 + \CA^2 C_c + \CF \CA D_{1;c} +\NF  \CF E_{1;c} 
 + \NF \CA E_{2;c} 
+  \NF^2 F_{1;c}  
\Biggr], \nonumber \\
\label{eq:fin2qqg}
\\
\Fintwo{\qq;\; c} &=& 2 N 
\Biggl[
\left( \sum_q Q_q^2\right) \TR \CF A_c
+ \CF^2 B_c  + \CF \CA D_{2;c}
+\NF  \CF E_{3;c}   
\Biggr],
\label{eq:fin2qq}
\end{eqnarray}
and
\begin{equation}
{\Fintwo{\ee}}_c = 
2 \Biggl[
\left( \sum_f Q_f^4\right)  A_c
+ B_c  
+\NF^\prime  E_{4;c}  
+  {\NF^\prime}^2 F_{2;c} 
\Biggr],
\label{eq:fin2ee}
\end{equation}
where the subscript $c= s, u$ denotes the channel. $Q_q$ refers to the
charge  of each of the $q$ active massless quark flavours for the QCD
processes and equivalently $Q_f$ denotes the charge of each of the $f$
active massless fermion  flavours  participating in the $\ee$ process.
We see that the ``light-by-light" contribution, $A_c$, due to 
internal fermion box
graphs and the ``abelian gluon" contribution, $B_c$, illustrated in
Fig.~\ref{fig:rep},
occur in all three processes.

The values of $A_c$, $B_{c}$, $C_{c}$, $D_{i;c}$, $E_{i;c}$ and $F_{i;c}$ are 
presented in Appendix~\ref{app:stwo} for the $s$-channel and in  
Appendix~\ref{app:utwo} for the $u$-channel.

\section{Finite one-loop contributions}
\label{sec:one}

In this section, we give explicit expressions for the finite remainder of 
the one-loop self-interference contributions $\Finone{\proc}$ defined as, 
\begin{equation}
 \Finone{\proc}(s,t,u) = \A_\proc^{\nnlo \, (1\times 1)}(s,t,u) 
- \Polone{\proc}(s,t,u).
\end{equation} 
Note again that the $\Polone{\proc}(s,t,u)$ is expanded through to
$\O{1}$ and  therefore contains finite as well as singular
contributions.   In fact, as in~\cite{ qqgg,1loopsquare, 1loopgggg}, all
polylogarithms are  collected in the singular part ($\Polone{\proc}$),
leaving finite remainders  with only logarithms and constants such as
$\pi^2$ and $\zeta_3$.  This is because in the expansion of the box
integral in $D=6$ dimensions, (see Appendix~\ref{app:master_int}) the
polylogarithms appear in  the $\O{\ep}$ and $\O{\ep^2}$ terms.  In order
for these polylogarithms to contribute at $\O{1}$, they must be
multiplied by an infrared singular term and are therefore contained in
$\Polone{\proc}$. The finite part of the one-loop self-interference
contribution due to the interference of one box graph with another only
collects the purely logarithmic $\O{1}$ terms in each.

Specifically we find, 
\begin{eqnarray}
\label{eq:fin1qqg}
\Finone{\qqg;\; c} &=& N \CF 
\Biggl[ \CF^2 G_{1;c} + \CA \CF G_{2;c} + \CA^2 G_{3;c}  \nonumber \\ 
&& \hspace{2cm}+ \NF \CF  X_{1;c} + \NF \CA X_{2;c} +  \NF^2 X_{3;c} \Biggr],\\
\Finone{\qq;\; c} &=& N \CF^2 G_{1;c} 
\label{eq:fin1qq}
\end{eqnarray}
and
\begin{equation}
\label{eq:fin1ee}
{\Finone{\ee}}_c = G_{1;c} + \NF^\prime X_{4;c} + {\NF^\prime}^2
  X_{5;c},
\end{equation}
where the subscript $c= s, u$ denotes the channel. 

The values of $G_{i;c}$ and $X_{i;c}$ are 
presented in Appendix~\ref{app:sone} for the $s$-channel and in  
Appendix~\ref{app:uone} for the $u$-channel.
 
\newpage
\section{Conclusions}
\label{sec:conc}

In this paper we calculated the NNLO QED and  QCD virtual corrections for
a range of $2 \to 2$ massless scattering  processes with external state
containing photons, namely $\qqg$, $\qq$, $\ee$  and  the processes
related by crossing symmetry and time reversal,  in the high energy
limit where the fermion masses can safely be ignored.  We used
conventional dimensional regularisation  and to remove the UV divergences
we renormalised using the \MSbar\ scheme. The renormalised amplitudes are
infrared divergent and contain poles  to $\O{1/\ep^4}$. In fact we were
able to check our results by comparing the infrared structure of the
calculated  amplitudes with the prediction of Catani's
formalism~\cite{catani} for the  infrared structure of generic one- and
two-loop QCD amplitudes.  Catani's method does not determine the  $1/\ep$
poles exactly, but suggests that the undetermined non-logarithmic
contribution $H^{(2)}$ can be extracted from a few basic two-loop
amplitudes. From the amplitudes  calculated in this paper together with
Refs.~\cite{qqgg, gggg, zqqg} we find  that, when summing over colours,
each external-leg contributes  independently to the leading order term in
$\ep$ of $H^{(2)}$.  The contributions of quarks, anti-quarks and gluons
and  their QED analogues are given in Section~\ref{sec:infra} for the
\MSbar\  scheme.

The main result of our paper is the finite remainders of the NNLO virtual
corrections after we subtract the pole structures of Catani's formalism
expanded through to $\O{1}$.  In Section~\ref{sec:two} we gave the finite
remainders of the interference  of the tree and the two-loop amplitude
and in Section~\ref{sec:one} we presented the finite remainders of the
self-interference of the one-loop  amplitude for each of the processes
under consideration.  The results are expressed in terms of
polylogarithms and  logarithms that are real in the physical region. The
renormalised amplitudes  for the three processes $\qqg$, $\qq$ and $\ee$,
differ with respect to  group invariants, the renormalisation procedure,
and their flavour content.  

The aim, of course, are more precise  predictions for the basic
scattering processes. Initial studies suggest that at NNLO an accuracy of
a few percent is attainable for strong interaction processes. However,
much work is needed to accomplish this goal.   An important ingredient is
a systematic procedure for analytically carrying through the cancellation
of the infrared singularities present in the virtual contributions
against the contributions from the one-loop $2 \to 3$ processes when one
particle is unresolved and  the tree-level $2 \to 4$ processes when two
particles are unresolved.  Such a method has not yet been established,
although the single and double unresolved limits of the relevant matrix
elements are well known~\cite{tc, ds,sone, cone}. In fact, many of the
analytic phase space  integrations for the double unresolved and single
unresolved loop contributions have already been studied in the context of
$e^+e^- \to $~photon + jet at ${\cal O}(\alpha \alpha_s)$~\cite{aude} and
Higgs production in hadron colliders~\cite{grazzini}.  A further
complication may be due to initial state radiation where the three loop
splitting functions~\cite{moms1,moms2,Gra1,NV,NVplb} are needed to extract
parton density functions~\cite{MRS} at an accuracy matching that of the
hard scattering matrix element. Nevertheless, it is an important task to
achieve more reliable theoretical calculations  that can take advantage
of the improving experimental data and  make a better test of the
underlying physics at short distances.

\section*{Acknowledgements}

C.A. thanks the theory group at Fermilab for their kind hospitality
during  the early stages of this work. C.A. would like to thank Zvi Bern
and Lance  Dixon for useful conversations.  We gratefully acknowledge the support of the British Council
and German Academic Exchange Service under ARC project 1050.  This work
was supported in part by the EU Fourth Framework Programme `Training and
Mobility of Researchers', Network `Quantum Chromodynamics and the Deep
Structure of Elementary Particles', contract FMRX-CT98-0194 (DG-12-MIHT)
and by the National Science Foundation, grant PHY9722101.
\newpage

\appendix

\section{Finite two-loop contributions}
\subsection{$s$-channel}
\label{app:stwo}
In this section we list the finite contributions for one-and two-loop
contributions in the $s$-channel as defined in
Eqs.~(\ref{eq:fin2qqg}), (\ref{eq:fin2qq}), (\ref{eq:fin2ee}), 
(\ref{eq:fin1qqg}),
(\ref{eq:fin1qq}) and
(\ref{eq:fin1ee}).

\begin{eqnarray}
{A_s}&=&{}\Biggl
[{128}\,{\Lip{4}{z}}-{128}\,{\Lip{4}{x}}+{128}\,{\Lip{4}{y}}+{}\Biggl
({}-{64\over 3}+{128}\,{\Ly}\Biggr ){}\,{\Lip{3}{x}} \nonumber \\
&&+{}\Biggl ({64\over
  3}\,{\Lx}-{64\over 3}\,{\pi^2}\Biggr ){}\,{\Lip{2}{x}}+{16\over 3}\,{\Lx^4}
-{64\over 3}\,{\Lx^3}\,{\Ly}+{}\Biggl
({}-{16}+{32\over 3}\,{\pi^2}+{32}\,{\Ly^2}\Biggr ){}\,{\Lx^2} \nonumber \\
&&+{}\Biggl
({}-{64\over 3}\,{\pi^2}\,{\Ly}+{48}+{160\over 9}\,{\pi^2}\Biggr ){}\,{\Lx}
+{64\over 3}\,{\zeta_3}+{224\over 45}\,{\pi^4}-{128}\,{\Ly}\,{\zeta_3}\Biggr
]{}\,{\tou} \nonumber \\
&&+{}\Biggl [{32\over 3}\,{\Lip{3}{x}}-{32\over 3}\,{\Lip{3}{y}}+{}\Biggl
({}-{32\over 3}\,{\Lx}-{32\over 3}\,{\Ly}\Biggr ){}\,{\Lip{2}{x}} \nonumber \\
&&+{}\Biggl ({}-{32\over 3}\,{\Ly^2}-{80\over 9}\,{\pi^2}-{64\over 3}\Biggr
){}\,{\Lx}+{}\Biggl ({64\over 3}+{32\over 3}\,{\pi^2}\Biggr ){}\,{\Ly}\Biggr
]{\ttoss}+{24}\,{\Lx^2}\,{\ttouu} \nonumber \\
&&+{}\Biggl [{416\over 3}\,{\Lip{3}{x}}+{64}\,{\Lip{3}{y}}\,{\Lx}-{416\over
  3}\,{\Lip{2}{x}}\,{\Lx}+{}\Biggl ({8}\,{\Ly^2}+{16}\Biggr ){}\,{\Lx^2} \nonumber \\
&&+{}\Biggl ({}-{8\over 3}\,{\Ly}+{80\over 3}+{112\over
  9}\,{\pi^2}-{64}\,{\zeta_3}-{64}\,{\Ly^2}\Biggr ){}\,{\Lx}-{416\over
  3}\,{\zeta_3}-{148\over 9}\,{\pi^2}+{44\over 45}\,{\pi^4}\Biggr ] 
\nonumber \\
& & + \Biggl\{t \leftrightarrow u \Biggr\} \\
{B_s}&=&{}\Biggl ({}-{112}\,{\Lip{4}{z}}-{88}\,{\Lip{4}{y}}+{}\Biggl
({}-{128}\,{\Ly}+{48}\,{\Lx}-{64}\Biggr ){}\,{\Lip{3}{x}} \nonumber \\
&&+{}\Biggl ({}-{16}\,{\Ly}-{16}\,{\Lx}+{12}\Biggr ){}\,{\Lip{3}{y}}+{}\Biggl
({12}\,{\Ly}-{4}\,{\Ly^2}+{8}\,{\Lx^2}-{8}\,{\pi^2}+{64}\,{\Lx}\Biggr
){}\,{\Lip{2}{x}} \nonumber \\
&& +{2\over 3}\,{\Lx^4}+{56\over 3}\,{\Lx^3}\,{\Ly}+{}\Biggl
({44}\,{\Ly}-{4}\,{\pi^2}+{2}-{32}\,{\Ly^2}\Biggr ){}\,{\Lx^2} \nonumber \\
&&+{}\Biggl ({}-{4}\,{\Ly^3}-{8}-{32}\,{\zeta_3}-{80\over
  3}\,{\pi^2}+{6}\,{\Ly^2}+{56\over 3}\,{\pi^2}\,{\Ly}\Biggr ){}\,{\Lx} \nonumber \\
&&+{\Ly^4}+{6}\,{\Ly^3}+{}\Biggl ({}-{10\over 3}\,{\pi^2}-{5}\Biggr
){}\,{\Ly^2}+{}\Biggl ({}-{39}-{18}\,{\pi^2}+{144}\,{\zeta_3}\Biggr
){}\,{\Ly} \nonumber \\
&&+{3}\,{\Ls}+{187\over 4}-{4}\,{\pi^2}\,{\Ls}+{4\over
  45}\,{\pi^4}-{5}\,{\pi^2}-{20}\,{\zeta_3}+{48}\,{\zeta_3}\,{\Ls}\Biggr
){}\,{\tou} \nonumber \\
&&+{}\Biggl ({}-{12}\,{\Lx^2}+{}\Biggl ({24}\,{\Ly}+{24}\Biggr
){}\,{\Lx}-{12}\,{\Ly^2}-{24}\,{\Ly}-{12}\,{\pi^2}\Biggr
){\ttoss}+{8}\,{\Lx^2}\,{\ttouu} \nonumber \\
&& +{}\Biggl ({}-{80}\,{\Lip{4}{y}}+{32}\,{\Lx}\,{\Lip{3}{x}}+{}\Biggl
({}-{128}\,{\Lx}-{152}\Biggr
){}\,{\Lip{3}{y}}+{152}\,{\Lip{2}{x}}\,{\Lx} \nonumber \\
&&+{8}\,{\Ly^2}\,{\Lip{2}{y}}+{}\Biggl ({}-{16}\,{\Ly^2}-{24}\Biggr ){}\,{\Lx^2}+{}\Biggl
({60}\,{\Ly^2}+{}\Biggl ({28}+{32\over 3}\,{\pi^2}\Biggr
){}\,{\Ly}-{58}\Biggr ){}\,{\Lx} \nonumber \\
&&+{14\over 3}\,{\Ly^4}+{44\over 3}\,{\Ly^3}+{8\over
  3}\,{\Ly^2}\,{\pi^2}+{}\Biggl ({96}\,{\zeta_3}-{32\over 3}\,{\pi^2}\Biggr
){}\,{\Ly}+{32\over 45}\,{\pi^4}+{16}\,{\zeta_3}-{86\over
  3}\,{\pi^2}-{2}\Biggr ) \nonumber \nonumber \\ & & {}\,{\one} \nonumber \\ 
& & + \Biggl\{t \leftrightarrow u \Biggr\} \\
{C_s}&=&{}\Biggl [{}-{20}\,{\Lip{4}{z}}+{28}\,{\Lip{4}{x}}+{}\Biggl
({}-{28}\,{\Ly}-{10}\,{\Lx}+{1\over 3}\Biggr ){}\,{\Lip{3}{x}} \nonumber \\
&&+{}\Biggl ({6}\,{\Lx^2}+{}\Biggl ({}-{1\over
  3}-{4}\,{\Ly}\Biggr ){}\,{\Lx}-{2}\,{\Ly^2}+{58\over 3}\,{\Ly}+{4\over
  3}\,{\pi^2}\Biggr ){}\,{\Lip{2}{x}} \nonumber \\
&&+{}\Biggl({58\over
  3}-{12}\,{\Lx}-{12}\,{\Ly}\Biggr){}\,{\Lip{3}{y}}-{1\over
  6}\,{\Lx^4}+{}\Biggl ({10\over 3}\,{\Ly}+{13\over 9}\Biggr ){}\,{\Lx^3} \nonumber \\
&&+{}\Biggl ({}-{9}\,{\Ly^2}-{1\over 3}\,{\pi^2}+{4\over 9}+{11\over
  2}\,{\Ls}+{13}\,{\Ly}\Biggr ){}\,{\Lx^2} \nonumber \\
&&+{}\Biggl ({}-{8\over 3}\,{\Ly^3}+{50\over 3}\,{\Ly^2}+{}\Biggl ({17\over
  3}\,{\pi^2}-{28\over 3}+{11}\,{\Ls}\Biggr ){}\,{\Ly}-{563\over
  27}-{233\over 36}\,{\pi^2}+{55\over 3}\,{\Ls}\Biggr ){}\,{\Lx} \nonumber \\
&&+{}\Biggl ({}-{2\over 3}\,{\pi^2}+{80\over 9}\Biggr ){}\,{\Ly^2}+{}\Biggl
({}-{284\over 27}-{299\over 36}\,{\pi^2}+{26}\,{\zeta_3}+{55\over
  3}\,{\Ls}\Biggr ){}\,{\Ly}-{209\over 36}\,{\pi^2}\,{\Ls} \nonumber \\
&&-{2}\,{\zeta_3}\,{\Ls}+{121\over 12}\,{\Ls^2}-{13}\,{\Ls}-{1142\over
  81}-{197\over 360}\,{\pi^4}+{461\over 36}\,{\pi^2}-{55\over
  18}\,{\zeta_3}\Biggr ]{}\,{\tou} \nonumber \\
&&+{}\Biggl [{}-{5\over 4}\,{\Lx^2}+{}\Biggl ({5\over 2}+{5\over
  2}\,{\Ly}\Biggr ){}\,{\Lx}-{5\over 4}\,{\Ly^2}-{5\over 2}\,{\Ly}-{5\over
  4}\,{\pi^2}\Biggr ]{}\,{\ttoss}+{1\over 2}\,{\Lx^2}\,{\ttouu} \nonumber \\
&&+{}\Biggl [{24}\,{\Lip{4}{y}}-{20}\,{\Lx}\,{\Lip{3}{x}}+{}\Biggl
({}-{40}\,{\Lx}-{22}\Biggr
){}\,{\Lip{3}{y}}+{22}\,{\Lip{2}{x}}\,{\Lx}+{8}\,{\Ly^2}\,{\Lip{2}{y}} \nonumber \\
&&
+{4\over 3}\,{\Lx^3}\,{\Ly}+{}\Biggl ({}-{6}\,{\Ly^2}-{575\over 36}\Biggr
){}\,{\Lx^2}+{}\Biggl ({46\over 3}\,{\Ly^2}+{}\Biggl ({73\over
  12}+{4}\,{\pi^2}\Biggr ){}\,{\Ly}-{637\over 18}\Biggr ){}\,{\Lx} \nonumber \\
&&+{1\over 3}\,{\Ly^4}+{59\over 9}\,{\Ly^3}+{}\Biggl ({2\over
  3}\,{\pi^2}+{11}\,{\Ls}\Biggr ){}\,{\Ly^2}+{}\Biggl
({44}\,{\zeta_3}-{4\over 9}\,{\pi^2}+{11}\,{\Ls}\Biggr ){}\,{\Ly} \nonumber \\
&& -{38\over 45}\,{\pi^4}+{77\over 72}\,{\pi^2}+{2}\,{\zeta_3}\Biggr ]
+ \Biggl\{t \leftrightarrow u \Biggr\} \\
{D_{1;s}}&=&{}\Biggl
[{96}\,{\Lip{4}{z}}-{48}\,{\Lip{4}{x}}+{52}\,{\Lip{4}{y}}+{}\Biggl
({124}\,{\Ly}-{8}\,{\Lx}+{46}\Biggr ){}\,{\Lip{3}{x}} \nonumber \\
&&+{}\Biggl
({}-{16}\,{\Lx^2}+{}\Biggl ({}-{46}+{8}\,{\Ly}\Biggr
){}\,{\Lx}+{6}\,{\Ly^2}-{30}\,{\Ly}+{4\over 3}\,{\pi^2}\Biggr
){}\,{\Lip{2}{x}} \nonumber \\
&&+{}\Biggl ({}-{30}+{28}\,{\Ly}+{36}\,{\Lx}\Biggr ){}\,{\Lip{3}{y}}+{1\over
  2}\,{\Lx^4}+{}\Biggl ({}-{56\over 3}\,{\Ly}-{100\over 9}\Biggr ){}\,{\Lx^3}
\nonumber \\
&&+{}\Biggl ({}-{125\over 3}\,{\Ly}+{39}\,{\Ly^2}+{214\over
  9}+{3}\,{\pi^2}-{22}\,{\Ls}\Biggr ){}\,{\Lx^2} \nonumber \\
&&+{}\Biggl ({14\over 3}\,{\Ly^3}-{73\over 3}\,{\Ly^2}+{}\Biggl
({}-{24}\,{\pi^2}+{4}\Biggr ){}\,{\Ly}+{155\over 9}\,{\pi^2}+{148\over
  3}\Biggr ){}\,{\Lx} \nonumber \\
&& -{74\over 9}\,{\Ly^3}+{}\Biggl ({10\over 3}\,{\pi^2}-{55\over
  9}-{11}\,{\Ls}\Biggr ){}\,{\Ly^2}+{}\Biggl ({}-{33}\,{\Ls}+{136\over 9}\,{\pi^2}-{140}\,{\zeta_3}+{241\over
  3}\Biggr ){}\,{\Ly} \nonumber \\
&&-{43417\over 324}+{23\over
  6}\,{\pi^2}\,{\Ls}-{52}\,{\zeta_3}\,{\Ls}-{173\over 18}\,{\pi^2}+{227\over
  180}\,{\pi^4}+{1834\over 27}\,{\Ls}+{515\over 9}\,{\zeta_3}\Biggr
]{}\,{\tou} \nonumber \\
&&+{}\Biggl [{14}\,{\Lx^2}+{}\Biggl ({}-{28}-{28}\,{\Ly}\Biggr
){}\,{\Lx}+{14}\,{\Ly^2}+{28}\,{\Ly}+{14}\,{\pi^2}\Biggr
]{}\,{\ttoss}-{5}\,{\Lx^2}\,{\ttouu} \nonumber \\
&&+{}\Biggl [{}-{8}\,{\Lip{4}{y}}+{24}\,{\Lx}\,{\Lip{3}{x}}+{}\Biggl
({144}\,{\Lx}+{120}\Biggr
){}\,{\Lip{3}{y}}-{120}\,{\Lip{2}{x}}\,{\Lx}-{20}\,{\Ly^2}\,{\Lip{2}{y}} \nonumber \\
&&-{8\over 3}\,{\Lx^3}\,{\Ly}+{}\Biggl ({20}\,{\Ly^2}+{472\over 9}\Biggr
){}\,{\Lx^2}+{}\Biggl ({}-{182\over 3}\,{\Ly^2}+{}\Biggl ({}-{40\over
  3}\,{\pi^2}-{104\over 3}\Biggr ){}\,{\Ly}+{898\over 9}\Biggr ){}\,{\Lx} \nonumber \\
&&-{3}\,{\Ly^4}-{184\over 9}\,{\Ly^3}+{}\Biggl ({}-{22}\,{\Ls}-{8\over
  3}\,{\pi^2}\Biggr ){}\,{\Ly^2}+{}\Biggl ({}-{22}\,{\Ls}+{56\over
  9}\,{\pi^2}-{136}\,{\zeta_3}\Biggr ){}\,{\Ly} \nonumber \\
&&+{4\over 3}\,{\pi^4}+{148\over 9}\,{\pi^2}+{2}-{12}\,{\zeta_3}\Biggr ]
+ \Biggl\{t \leftrightarrow u \Biggr\} \\
{E_{1;s}}&=&{}\Biggl [{22\over 9}\,{\Lx^3}+{}\Biggl ({}-{76\over
  9}+{4}\,{\Ls}+{2\over 3}\,{\Ly}\Biggr ){}\,{\Lx^2}+{}\Biggl ({1\over
  3}\,{\Ly^2}+{\Ly}+{16\over 9}\,{\pi^2}-{7\over 3}\Biggr ){}\,{\Lx} \nonumber \\
&&+{11\over 9}\,{\Ly^3}+{}\Biggl ({7\over 9}+{2}\,{\Ls}\Biggr
){}\,{\Ly^2}+{}\Biggl ({6}\,{\Ls}+{8\over 9}\,{\pi^2}-{37\over 3}\Biggr
){}\,{\Ly} \nonumber \\
&&+{19\over 9}\,{\pi^2}-{328\over 27}\,{\Ls}+{3401\over 162}-{2\over
  9}\,{\zeta_3}-{1\over 3}\,{\pi^2}\,{\Ls}\Biggr ]{}\,{\tou} \nonumber \\
&&+{}\Biggl [{}-{46\over 9}\,{\Lx^2}+{}\Biggl ({2\over 3}\,{\Ly}+{2\over
  3}\,{\Ly^2}-{76\over 9}\Biggr ){}\,{\Lx} \nonumber \\
&&+{22\over 9}\,{\Ly^3}+{4}\,{\Ly^2}\,{\Ls}+{}\Biggl ({16\over
  9}\,{\pi^2}+{4}\,{\Ls}\Biggr ){\Ly}+{8\over 9}\,{\pi^2}\Biggr ]+ \Biggl\{t \leftrightarrow u \Biggr\} \\
{E_{2;s}}&=&{}\Biggl [{}-{4\over 3}\,{\Lip{3}{x}}-{4\over
  3}\,{\Lip{3}{y}}+{}\Biggl ({}-{4\over 3}\,{\Ly}+{4\over 3}\,{\Lx}\Biggr
){}\,{\Lip{2}{x}} \nonumber \\
&&-{11\over 18}\,{\Lx^3}+{}\Biggl ({}-{1\over 2}\,{\Ly}+{5\over
  6}-{\Ls}\Biggr ){}\,{\Lx^2} \nonumber \\
&&+{}\Biggl ({}-{5\over 3}\,{\Ly^2}+{}\Biggl ({10\over 9}-{2}\,{\Ls}\Biggr
){}\,{\Ly}-{1\over 9}\,{\pi^2}+{37\over 9}-{31\over 6}\,{\Ls}\Biggr
){}\,{\Lx} \nonumber \\
&&-{41\over 18}\,{\Ly^2}+{}\Biggl ({5\over 9}\,{\pi^2}+{43\over 9}-{31\over
  6}\,{\Ls}\Biggr ){}\,{\Ly}+{65\over 81}+{19\over
  18}\,{\pi^2}\,{\Ls}-{11\over 3}\,{\Ls^2} \nonumber \\
&&-{13\over 9}\,{\zeta_3}+{206\over
  27}\,{\Ls}-{275\over 108}\,{\pi^2}\Biggr ]{}\,{\tou} \nonumber \\
&&+{}\Biggl [{}-{\Lx^2}+{}\Biggl ({2}\,{\Ly}+{2}\Biggr
){}\,{\Lx}-{\Ly^2}-{2}\,{\Ly}-{\pi^2}\Biggr ]{}\,{\ttoss} \nonumber \\
&&+{}\Biggl [{14\over 9}\,{\Lx^2}+{}\Biggl ({2\over 3}\,{\Ly}-{1\over
  3}\,{\Ly^2}+{38\over 9}\Biggr ){}\,{\Lx}-{11\over
  9}\,{\Ly^3}-{2}\,{\Ly^2}\,{\Ls} \nonumber \\
&&-{17\over 18}\,{\pi^2}+{}\Biggl ({}-{8\over
  9}\,{\pi^2}-{2}\,{\Ls}\Biggr ){}\,{\Ly}\Biggr ] + \Biggl\{t \leftrightarrow u \Biggr\} \\
{F_{1;s}}&=&{}\Biggl [{5\over 36}\,{\Lx^2}+{}\Biggl ({}-{10\over 27}+{1\over
  3}\,{\Ls}+{1\over 18}\,{\Ly}\Biggr ){}\,{\Lx}+{5\over 36}\,{\Ly^2}+{}\Biggl
({}-{10\over 27}+{1\over 3}\,{\Ls}\Biggr ){}\,{\Ly} \nonumber \\
&&+{1\over 3}\,{\Ls^2}+{1\over 54}\,{\pi^2}-{20\over 27}\,{\Ls}\Biggr
]{}\,{\tou} + \Biggl\{t \leftrightarrow u \Biggr\} \\
{D_{2;s}}&=&{}\Biggl
[{48}\,{\Lip{4}{z}}-{16}\,{\Lip{4}{x}}+{24}\,{\Lip{4}{y}}+{}\Biggl
({56}\,{\Ly}-{8}\,{\Lx}+{20}\Biggr ){}\,{\Lip{3}{x}} \nonumber \\
&&+{}\Biggl ({8}\,{\Lx}-{12}+{16}\,{\Ly}\Biggr ){}\,{\Lip{3}{y}}+{}\Biggl
({16\over 3}\,{\pi^2}-{20}\,{\Lx}-{12}\,{\Ly}-{8}\,{\Lx^2}+{4}\,{\Ly^2}\Biggr
){}\,{\Lip{2}{x}} \nonumber \\
&&+{1\over 3}\,{\Lx^4}+{}\Biggl ({}-{8}\,{\Ly}-{70\over 9}\Biggr
){}\,{\Lx^3}+{}\Biggl ({}-{4}\,{\pi^2}+{286\over
  9}-{16}\,{\Ly}+{14}\,{\Ly^2}-{44\over 3}\,{\Ls}\Biggr ){}\,{\Lx^2} \nonumber \\
&&+{}\Biggl ({}-{22\over
  9}\,{\pi^2}+{4}\,{\Ly^3}-{8}\,{\pi^2}\,{\Ly}-{6}\,{\Ly^2}\Biggr
){}\,{\Lx}-{44\over 9}\,{\Ly^3}+{}\Biggl ({}-{4\over 3}\,{\pi^2}+{35\over
  9}-{22\over 3}\,{\Ls}\Biggr ){}\,{\Ly^2} \nonumber \\
&&+{}\Biggl ({57}-{26\over 9}\,{\pi^2}-{72}\,{\zeta_3}-{22}\,{\Ls}\Biggr
){}\,{\Ly}+{479\over 9}\,{\zeta_3}+{19\over
  60}\,{\pi^4}-{52}\,{\zeta_3}\,{\Ls}+{1141\over 27}\,{\Ls}-{215\over
  18}\,{\pi^2} \nonumber \\
&&-{43417\over 324}+{23\over 6}\,{\pi^2}\,{\Ls}\Biggr ]{}\,{\tou}
\nonumber \\
&&+{}\Biggl [{6}\,{\Lx^2}+{}\Biggl ({}-{12}-{12}\,{\Ly}\Biggr
){}\,{\Lx}+{6}\,{\Ly^2}+{12}\,{\Ly}+{6}\,{\pi^2}\Biggr
]{}\,{\ttoss}-{6}\,{\Lx^2}\,{\ttouu} \nonumber \\
&&+{}\Biggl
[{16}\,{\Lip{4}{y}}+{48}\,{\Lip{3}{x}}\,{\Ly}+{64}\,{\Lip{3}{y}}-{8}\,{\Ly^2}\,{\Lip{2}{y}}-{64}\,{\Lip{2}{x}}\,{\Lx}
\nonumber \\
&&-{4\over 3}\,{\Lx^4}+{}\Biggl ({}-{20\over 3}\,{\pi^2}+{6}\,{\Ly^2}\Biggr
){}\,{\Lx^2}+{}\Biggl ({}-{24}\,{\Ly^2}+{}\Biggl ({}-{16\over
  3}\,{\pi^2}-{14}\Biggr ){}\,{\Ly}-{148\over 9}\,{\pi^2}\Biggr ){}\,{\Lx} \nonumber \\
&&-{112\over 9}\,{\Ly^3}+{}\Biggl ({}-{44\over 3}\,{\Ls}+{298\over 9}\Biggr
){}\,{\Ly^2}+{}\Biggl ({538\over 9}-{48}\,{\zeta_3}-{44\over 3}\,{\Ls}\Biggr
){}\,{\Ly}-{8}\,{\zeta_3}-{1\over 3}\,{\pi^4}+{61\over 9}\,{\pi^2}\Biggr ]
\nonumber \\
& & + \Biggl\{t \leftrightarrow u \Biggr\} \\
{E_{3;s}}&=&{}\Biggl [{16\over 9}\,{\Lx^3}+{}\Biggl ({}-{76\over 9}+{8\over
  3}\,{\Ls}\Biggr ){}\,{\Lx^2}+{16\over 9}\,{\pi^2}\,{\Lx} \nonumber \\
&&+{8\over 9}\,{\Ly^3}+{}\Biggl ({4\over 3}\,{\Ls}-{2\over 9}\Biggr
){}\,{\Ly^2}+{}\Biggl ({8\over 9}\,{\pi^2}+{4}\,{\Ls}-{10}\Biggr
){}\,{\Ly} \nonumber \\
&&-{1\over 3}\,{\pi^2}\,{\Ls}-{202\over 27}\,{\Ls}+{19\over
  9}\,{\pi^2}-{2\over 9}\,{\zeta_3}+{3401\over 162}\Biggr ]{}\,{\tou} \nonumber \\
&&+{}\Biggl [{16\over 9}\,{\pi^2}\,{\Lx}+{16\over 9}\,{\Ly^3}+{}\Biggl
({8\over 3}\,{\Ls}-{52\over 9}\Biggr ){}\,{\Ly^2} \nonumber \\
&&+{}\Biggl ({}-{76\over 9}+{8\over 3}\,{\Ls}\Biggr ){}\,{\Ly}+{8\over
  9}\,{\pi^2}\Biggr ]+ \Biggl\{t \leftrightarrow u \Biggr\} \\
{E_{4;s}}&=&{}\Biggl [{16\over 9}\,{\Ly^3}-{4\over 9}\,{\Ly^2}+{16\over
  3}\,{\Ly^2}\,{\Ls}-{20}\,{\Ly}+{64\over
  9}\,{\pi^2}\,{\Ly}+{16}\,{\Ly}\,{\Ls} \nonumber \\
&&+{32\over 9}\,{\Lx^3}-{152\over
  9}\,{\Lx^2}+{32\over 3}\,{\Lx^2}\,{\Ls}+{128\over
  9}\,{\pi^2}\,{\Lx} \nonumber \\
&&-{2\over 3}\,{\pi^2}\,{\Ls}+{110\over
  9}\,{\pi^2}+{3401\over 81}-{908\over 27}\,{\Ls}-{4\over 9}\,{\zeta_3}\Biggr
]{}\,{\tou} \nonumber \\
&&+{}\Biggl [{32\over 9}\,{\Lx^3}-{104\over 9}\,{\Lx^2}+{32\over
  3}\,{\Lx^2}\,{\Ls}-{152\over 9}\,{\Lx}+{32\over 3}\,{\Lx}\,{\Ls} \nonumber \\
&&+{128\over 9}\,{\pi^2}\,{\Lx}+{64\over 9}\,{\pi^2}\Biggr ] + \Biggl\{t \leftrightarrow u \Biggr\} \\
{F_{2;s}}&=&{}\Biggl [{}-{92\over 27}\,{\pi^2}+{32\over 9}\,{\Ls^2}
-{160\over 27}\,{\Ls}\Biggr ]{}\,{\tou} + \Biggl\{t \leftrightarrow u \Biggr\}
\end{eqnarray}

\subsection{$u$-channel}
\label{app:utwo}
In this section we list the finite contributions for one-and two-loop
contributions in the $u$-channel as defined in
Eqs.~(\ref{eq:fin2qqg}), (\ref{eq:fin2qq}), (\ref{eq:fin2ee}), 
(\ref{eq:fin1qqg}),
(\ref{eq:fin1qq}) and
(\ref{eq:fin1ee}).

\begin{eqnarray}
{A_u}&=&{}\Biggl
[{24}\,{\pi^2}-{48}\,{\Lx}\,{\Ly}+{24}\,{\Ly^2}+{24}\,{\Lx^2}\Biggr
]{}\,{\ttoss} +{24}\,{\Ly^2}\,{\ssott} \nonumber \\
&&+{}\Biggl [{}\Biggl ({64}\,{\Ly}+{32\over 3}\Biggr )
{}\,{\Lip{3}{x}}+{64}\,{\Lip{3}{y}}\,{\Lx}-{32\over
  3}\,{\Lip{2}{x}}\,{\Lx}+{}\Biggl ({}-{8}+{16}\,{\Ly^2}\Biggr )
{}\,{\Lx^2} \nonumber \\
&& +{}\Biggl ({}\Biggl ({}-{64\over 3}\,{\pi^2}+{16}\Biggr
){}\,{\Ly}-{16\over 9}\,{\pi^2}+{24}-{64}\,{\zeta_3}+{32\over
  3}\,{\Ly^3}-{32\over 3}\,{\Ly^2}\Biggr ){}\,{\Lx}+{64\over 9}\,{\Ly^3} \nonumber \\
&&+{}\Biggl ({}-{48}+{64}\,{\zeta_3}+{32\over 9}\,{\pi^2}\Biggr
){}\,{\Ly}+{}\Biggl ({}-{16\over 3}\,{\pi^2}-{16}\Biggr ){}\,{\Ly^2}-{16\over
  3}\,{\Ly^4}-{8}\,{\pi^2}+{32\over 3}\,{\zeta_3} \nonumber \\
&&+{88\over 45}\,{\pi^4}\Biggr ]{}\,{\tpsost} \nonumber \\
&&+{}\Biggl
[{}-{128}\,{\Lip{4}{x}}-{128}\,{\Lip{4}{y}}+{128}\,{\Lip{4}{z}}+{}\Biggl
({64}\,{\Ly}+{32\over 3}\Biggr ){}\,{\Lip{3}{x}} \nonumber \\
&&+{}\Biggl ({128}\,{\Ly}+{64\over 3}-{64}\,{\Lx}\Biggr
){}\,{\Lip{3}{y}}+{}\Biggl ({}-{32\over 3}\,{\Lx}+{64\over 3}\,{\Ly}+{64\over
  3}\,{\pi^2}\Biggr ){}\,{\Lip{2}{x}} \nonumber \\
&&+{16\over 3}\,{\Lx^4}-{64\over 3}\,{\Lx^3}\,{\Ly}+{}\Biggl
({16}\,{\Ly^2}-{8}+{32\over 3}\,{\pi^2}\Biggr ){}\,{\Lx^2} \nonumber \\
&&+{}\Biggl ({32\over 3}\,{\Ly^2}+{32\over
  3}\,{\Ly^3}+{64}\,{\zeta_3}-{16\over 9}\,{\pi^2}+{24}+{16}\,{\Ly}\Biggr
){}\,{\Lx}+{88\over 15}\,{\pi^4}-{32\over 3}\,{\zeta_3} \nonumber \\
&&+{}\Biggl ({}-{32\over 9}\,{\pi^2}-{64}\,{\zeta_3}\Biggr
){}\,{\Ly}+{16}\,{\Ly^2}\,{\pi^2}-{8}\,{\pi^2}\Biggr ]{}\,{\tmsost} \nonumber \\
&&+{}\Biggl [{}-{32\over 3}\,{\Lip{3}{x}}-{64\over 3}\,{\Lip{3}{y}}+{}\Biggl
({32\over 3}\,{\Lx}-{64\over 3}\,{\Ly}\Biggr ){}\,{\Lip{2}{x}} \nonumber \\
&&+{}\Biggl ({}-{32\over 3}\,{\Ly^2}-{64\over 3}+{16\over 9}\,{\pi^2}\Biggr
){}\,{\Lx}+{32\over 9}\,{\pi^2}\,{\Ly}+{32\over 3}\,{\zeta_3}\Biggr
]{}\,{\tmsou} \nonumber \\
&&+{}\Biggl [{}\Biggl ({64}\,{\Ly}-{416\over 3}\Biggr
){}\,{\Lip{3}{x}}+{64}\,{\Lip{3}{y}}\,{\Lx}+{416\over
  3}\,{\Lip{2}{x}}\,{\Lx}+{}\Biggl ({64}\,{\Ly}+{16}+{16}\,{\Ly^2}\Biggr
){}\,{\Lx^2} \nonumber \\
&&+{}\Biggl ({}-{160\over 3}\,{\Ly^2}+{32\over 3}\,{\Ly^3}+{}\Biggl
({}-{80\over 3}-{64\over 3}\,{\pi^2}\Biggr ){}\,{\Ly}+{208\over
  9}\,{\pi^2}-{64}\,{\zeta_3}+{80\over 3}\Biggr ){}\,{\Lx}+{320\over
  9}\,{\Ly^3} \nonumber \\
&&+{}\Biggl ({}-{160\over 3}+{160\over 9}\,{\pi^2}+{64}\,{\zeta_3}\Biggr
){}\,{\Ly}+{}\Biggl ({}-{16\over 3}\,{\pi^2}+{80\over 3}\Biggr
){}\,{\Ly^2}-{16\over 3}\,{\Ly^4}+{88\over 45}\,{\pi^4}-{200\over 9}\,{\pi^2}
\nonumber \\
&&-{416\over 3}\,{\zeta_3}\Biggr ]{}\,{\one} \\
{B_u}&=&{}-{12}\,{\Lx^2}\,{\tpsou}+{24}\,{\Lx}\,{\tmsou} \nonumber \\
&&+{8}\,{\Ly^2}\,{\ssott}+{}\Biggl [{}-{16}\,{\Lx}\,{\Ly}+{8}\,{\Ly^2}+{8}\,{\pi^2}+{8}\,{\Lx^2}\Biggr ]{}\,{\ttoss}\nonumber \\
&&+{}\Biggl [{44}\,{\Lip{4}{z}}+{44}\,{\Lip{4}{y}}+{}\Biggl
({}-{16}\,{\Lx}-{56}\,{\Ly}+{26}\Biggr
){}\,{\Lip{3}{x}}-{88}\,{\Lip{3}{y}}\,{\Lx} \nonumber \\
&&+{}\Biggl ({}-{6}\,{\Lx^2}+{}\Biggl ({12}\,{\Ly}-{26}\Biggr
){}\,{\Lx}-{6}\,{\pi^2}\Biggr ){}\,{\Lip{2}{x}}+{5}\,{\Lx^4}+{}\Biggl
({}-{20}\,{\Ly}+{3}\Biggr ){}\,{\Lx^3} \nonumber \\
&&+{}\Biggl ({}-{3\over 2}+{10\over 3}\,{\pi^2}-{28}\,{\Ly}+{\Ly^2}\Biggr
){}\,{\Lx^2}+{}\Biggl ({}-{28\over 3}\,{\Ly^3}+{40}\,{\Ly^2}+{}\Biggl
({20\over 3}\,{\pi^2}+{3}\Biggr ){}\,{\Ly} \nonumber \\
&&-{1\over 3}\,{\pi^2}-{47\over 2}+{72}\,{\zeta_3}\Biggr ){}\,{\Lx}+{14\over
  3}\,{\Ly^4}-{80\over 3}\,{\Ly^3}+{}\Biggl ({}-{3}-{10\over
  3}\,{\pi^2}\Biggr ){}\,{\Ly^2} \nonumber \\
&&+{}\Biggl ({47}-{56}\,{\zeta_3}-{28\over 3}\,{\pi^2}\Biggr
){}\,{\Ly}-{13\over 2}\,{\pi^2}-{46}\,{\zeta_3}-{4}\,{\pi^2}\,{\LU}-{28\over
  9}\,{\pi^4} \nonumber \\
&&+{3}\,{\LU}+{48}\,{\zeta_3}\,{\LU}+{187\over 4}\Biggr ]{}\,{\tpsost} \nonumber \\
&&+{}\Biggl
[{}-{44}\,{\Lip{4}{z}}+{44}\,{\Lip{4}{y}}+{112}\,{\Lip{4}{x}} \nonumber \\
&&+{}\Biggl
({}-{32}\,{\Lx}+{38}-{24}\,{\Ly}\Biggr ){}\,{\Lip{3}{x}}+{}\Biggl ({24}\,{\Lx}-{48}\,{\Ly}+{76}\Biggr ){}\,{\Lip{3}{y}} \nonumber \\
&&+{}\Biggl ({}-{2}\,{\Lx^2}+{}\Biggl ({}-{38}+{4}\,{\Ly}\Biggr
){}\,{\Lx}+{76}\,{\Ly}-{4}\,{\Ly^2}+{6}\,{\pi^2}\Biggr ){}\,{\Lip{2}{x}} \nonumber \\
&&+{1\over 3}\,{\Lx^4}+{}\Biggl ({}-{4\over 3}\,{\Ly}-{3}\Biggr
){}\,{\Lx^3}+{}\Biggl ({}-{6}\,{\pi^2}-{\Ly^2}+{7\over 2}-{16}\,{\Ly}\Biggr
){}\,{\Lx^2} \nonumber \\
&&+{}\Biggl ({}-{8}\,{\Ly^3}+{54}\,{\Ly^2}+{}\Biggl ({12}\,{\pi^2}-{7}\Biggr
){}\,{\Ly}-{7\over 3}\,{\pi^2}-{56}\,{\zeta_3}+{31\over 2}\Biggr ){}\,{\Lx}
\nonumber \\
&&-{4\over 3}\,{\Ly^2}\,{\pi^2}+{}\Biggl ({24}\,{\zeta_3}-{86\over
  3}\,{\pi^2}\Biggr ){}\,{\Ly}-{206\over 45}\,{\pi^4}-{38}\,{\zeta_3}+{7\over
  2}\,{\pi^2}\Biggr ]{}\,{\tmsost} \nonumber \\
&&+{}\Biggl [{80}\,{\Lip{4}{z}}+{80}\,{\Lip{4}{y}}+{}\Biggl
({}-{96}\,{\Ly}-{32}\,{\Lx}+{152}\Biggr
){}\,{\Lip{3}{x}}-{160}\,{\Lip{3}{y}}\,{\Lx} \nonumber \\
&&+{}\Biggl ({}-{8}\,{\Lx^2}+{}\Biggl ({}-{152}+{16}\,{\Ly}\Biggr
){}\,{\Lx}-{8}\,{\pi^2}\Biggr ){}\,{\Lip{2}{x}}+{8}\,{\Lx^4}+{}\Biggl
({44\over 3}-{32}\,{\Ly}\Biggr ){}\,{\Lx^3} \nonumber \\
&&+{}\Biggl ({16\over 3}\,{\pi^2}-{4}\,{\Ly^2}-{104}\,{\Ly}-{24}\Biggr
){}\,{\Lx^2}+{}\Biggl ({}-{16}\,{\Ly^3}+{72}\,{\Ly^2}+{}\Biggl
({16}\,{\pi^2}-{8}\Biggr ){}\,{\Ly} \nonumber \\
&&+{4\over 3}\,{\pi^2}+{128}\,{\zeta_3}-{58}\Biggr
){}\,{\Lx}+{8}\,{\Ly^4}-{48}\,{\Ly^3}+{}\Biggl ({8}-{8\over 3}\,{\pi^2}\Biggr
){}\,{\Ly^2} \nonumber \\
&&+{}\Biggl ({116}-{56\over 3}\,{\pi^2}-{96}\,{\zeta_3}\Biggr
){}\,{\Ly}-{4}-{76\over 3}\,{\pi^2}-{120}\,{\zeta_3}-{24\over
  5}\,{\pi^4}\Biggr ]{}\,{\one} \\
{C_u}&=&{}\Biggl [{}-{\Lx}\,{\Ly}+{1\over 2}\,{\pi^2}+{1\over
  2}\,{\Ly^2}+{1\over 2}\,{\Lx^2}\Biggr ]{}\,{\ttoss} +{1\over
  2}\,{\Ly^2}\,{\ssott} \nonumber \\
&&-{5\over 4}\,{\Lx^2}\,{\tpsou}+{5\over 2}\,{\Lx}\,{\tmsou} \nonumber \\
&&+{}\Biggl [{}-{14}\,{\Lip{4}{y}}-{14}\,{\Lip{4}{z}}+{}\Biggl ({}-{59\over
  6}+{11}\,{\Lx}-{31}\,{\Ly}\Biggr ){}\,{\Lip{3}{x}}-{9}\,{\Lip{3}{y}}\,{\Lx}\nonumber \\
&&+{}\Biggl ({}-{4}\,{\Lx^2}+{}\Biggl ({59\over 6}+{8}\,{\Ly}\Biggr
){}\,{\Lx}-{4}\,{\pi^2}\Biggr ){}\,{\Lip{2}{x}}-{1\over 4}\,{\Lx^4}+{}\Biggl
({4\over 3}\,{\Ly}+{13\over 18}\Biggr ){}\,{\Lx^3} \nonumber \\
&&+{}\Biggl ({}-{7\over 2}\,{\Ly^2}-{22\over 3}\,{\Ly}-{5\over
  2}\,{\pi^2}+{11\over 4}\,{\LU}+{14\over 3}\Biggr ){}\,{\Lx^2} \nonumber \\
&&+{}\Biggl ({}-{4\over 3}\,{\Ly^3}+{55\over 2}\,{\Ly^2}+{}\Biggl ({35\over
  6}\,{\pi^2}-{33\over 2}\,{\LU}\Biggr ){}\,{\Ly}+{2}\,{\zeta_3}+{55\over
  3}\,{\LU}-{847\over 54}+{1\over 18}\,{\pi^2}\Biggr ){}\,{\Lx} \nonumber \\
&&+{2\over 3}\,{\Ly^4}-{55\over 3}\,{\Ly^3}+{}\Biggl ({}-{1\over
  2}\,{\pi^2}+{33\over 2}\,{\LU}\Biggr ){}\,{\Ly^2}+{}\Biggl ({847\over
  27}-{28\over 9}\,{\pi^2}+{5}\,{\zeta_3}-{110\over 3}\,{\LU}\Biggr
){}\,{\Ly} \nonumber \\
&&-{1142\over 81}+{61\over 9}\,{\zeta_3}-{13}\,{\LU}-{166\over
  9}\,{\pi^2}+{47\over 360}\,{\pi^4}-{2}\,{\zeta_3}\,{\LU}+{143\over
  18}\,{\pi^2}\,{\LU}+{121\over 12}\,{\LU^2}\Biggr ]{}\,{\tpsost} \nonumber \\
&&+{}\Biggl
[{20}\,{\Lip{4}{x}}+{14}\,{\Lip{4}{y}}-{14}\,{\Lip{4}{z}}+{}\Biggl
({}-{7}\,{\Ly}+{19\over 2}-{\Lx}\Biggr ){}\,{\Lip{3}{x}} \nonumber \\
&&+{}\Biggl ({}-{14}\,{\Ly}+{7}\,{\Lx}+{19}\Biggr ){}\,{\Lip{3}{y}}+{}\Biggl
({}-{2}\,{\Lx^2}-{19\over 2}\,{\Lx}+{2\over 3}\,{\pi^2}+{19}\,{\Ly}\Biggr
){}\,{\Lip{2}{x}} \nonumber \\
&&-{1\over 4}\,{\Lx^4}+{}\Biggl ({2\over 3}\,{\Ly}+{13\over 18}\Biggr
){}\,{\Lx^3}+{}\Biggl ({}-{3\over 2}\,{\Ly^2}-{38\over 9}-{3\over
  2}\,{\pi^2}-{10}\,{\Ly}+{11\over 4}\,{\LU}\Biggr ){}\,{\Lx^2} \nonumber \\
&&+{}\Biggl ({}-{4\over 3}\,{\Ly^3}+{39\over 2}\,{\Ly^2}+{}\Biggl ({76\over
  9}+{13\over 6}\,{\pi^2}-{11\over 2}\,{\LU}\Biggr ){}\,{\Ly}+{77\over
  36}\,{\pi^2}-{31\over 6}-{12}\,{\zeta_3}\Biggr ){}\,{\Lx} \nonumber \\
&&-{3\over 2}\,{\Ly^2}\,{\pi^2}+{}\Biggl ({7}\,{\zeta_3}-{79\over
  6}\,{\pi^2}\Biggr ){}\,{\Ly}-{19\over 2}\,{\zeta_3}-{38\over
  9}\,{\pi^2}-{5\over 6}\,{\pi^4}+{11\over 4}\,{\pi^2}\,{\LU}\Biggr
]{}\,{\tmsost} \nonumber \\
&&+{}\Biggl [{}-{24}\,{\Lip{4}{y}}-{24}\,{\Lip{4}{z}}+{}\Biggl
({22}-{60}\,{\Ly}+{20}\,{\Lx}\Biggr
){}\,{\Lip{3}{x}}-{20}\,{\Lip{3}{y}}\,{\Lx} \nonumber \\
&&+{}\Biggl ({}-{8}\,{\Lx^2}+{}\Biggl ({}-{22}+{16}\,{\Ly}\Biggr
){}\,{\Lx}-{8}\,{\pi^2}\Biggr ){}\,{\Lip{2}{x}}-{2\over 3}\,{\Lx^4}+{}\Biggl
({4\over 3}\,{\Ly}+{59\over 9}\Biggr ){}\,{\Lx^3} \nonumber \\
&&+{}\Biggl ({}-{35}\,{\Ly}-{4\over 3}\,{\pi^2}-{575\over
  36}-{6}\,{\Ly^2}+{11}\,{\LU}\Biggr ){}\,{\Lx^2} \nonumber \\
&&+{}\Biggl ({}-{8\over 3}\,{\Ly^3}+{131\over 3}\,{\Ly^2}+{}\Biggl ({26\over
  3}\,{\pi^2}-{22}\,{\LU}+{178\over 9}\Biggr ){}\,{\Ly}+{11}\,{\LU}-{637\over
  18}+{24}\,{\zeta_3}+{53\over 9}\,{\pi^2}\Biggr ){}\,{\Lx} \nonumber \\
&&+{4\over 3}\,{\Ly^4}-{262\over 9}\,{\Ly^3}+{}\Biggl ({}-{178\over
  9}+{2\over 3}\,{\pi^2}+{22}\,{\LU}\Biggr ){}\,{\Ly^2}+{}\Biggl ({637\over
  9}-{28}\,{\zeta_3}-{67\over 9}\,{\pi^2}-{22}\,{\LU}\Biggr ){}\,{\Ly} \nonumber \\
&&-{18}\,{\zeta_3}+{2\over 45}\,{\pi^4}+{11}\,{\pi^2}\,{\LU}-{71\over
  3}\,{\pi^2}\Biggr ]{}\,{\one} \\
{D_{1;u}}&=&{}\Biggl
[{10}\,{\Lx}\,{\Ly}-{5}\,{\pi^2}-{5}\,{\Ly^2}-{5}\,{\Lx^2}\Biggr
]{}\,{\ttoss} -{5}\,{\Ly^2}\,{\ssott} \nonumber \\
&&+{14}\,{\Lx^2}\,{\tpsou}-{28}\,{\Lx}\,{\tmsou} \nonumber \\
&&+{}\Biggl [{}-{2}\,{\Lip{4}{y}}-{2}\,{\Lip{4}{z}}+{}\Biggl
({}-{8}-{10}\,{\Lx}+{90}\,{\Ly}\Biggr
){}\,{\Lip{3}{x}}+{70}\,{\Lip{3}{y}}\,{\Lx} \nonumber \\
&&+{}\Biggl ({11}\,{\Lx^2}+{}\Biggl ({8}-{22}\,{\Ly}\Biggr
){}\,{\Lx}+{11}\,{\pi^2}\Biggr ){}\,{\Lip{2}{x}}-{11\over
  6}\,{\Lx^4}+{}\Biggl ({28\over 3}\,{\Ly}-{29\over 3}\Biggr ){}\,{\Lx^3} \nonumber \\
&&+{}\Biggl ({}-{33\over 2}\,{\LU}+{53\over 6}+{47}\,{\Ly}+{13\over
  2}\,{\Ly^2}\Biggr ){}\,{\Lx^2}+{}\Biggl ({22\over
  3}\,{\Ly^3}-{75}\,{\Ly^2}+{}\Biggl ({}-{65\over
  3}-{15}\,{\pi^2}+{33}\,{\LU}\Biggr ){}\,{\Ly} \nonumber \\
&&+{389\over 6}-{60}\,{\zeta_3}-{31\over 3}\,{\pi^2}-{33\over 2}\,{\LU}\Biggr
){}\,{\Lx}-{11\over 3}\,{\Ly^4}+{50}\,{\Ly^3}+{}\Biggl ({8\over
  3}\,{\pi^2}+{65\over 3}-{33}\,{\LU}\Biggr ){}\,{\Ly^2} \nonumber \\
&&+{}\Biggl ({50}\,{\zeta_3}+{29\over 3}\,{\pi^2}-{389\over
  3}+{33}\,{\LU}\Biggr ){}\,{\Ly}+{587\over 9}\,{\zeta_3}+{326\over
  9}\,{\pi^2}-{52}\,{\zeta_3}\,{\LU}+{61\over 36}\,{\pi^4} \nonumber \\
&&+{1834\over 27}\,{\LU}-{43417\over 324}-{38\over 3}\,{\pi^2}\,{\LU}\Biggr
]{}\,{\tpsost} \nonumber \\
&&+{}\Biggl
[{}-{96}\,{\Lip{4}{x}}-{50}\,{\Lip{4}{y}}+{50}\,{\Lip{4}{z}}+{}\Biggl
({18}\,{\Lx}+{26}\,{\Ly}-{38}\Biggr ){}\,{\Lip{3}{x}} \nonumber \\
&&+{}\Biggl ({52}\,{\Ly}-{26}\,{\Lx}-{76}\Biggr ){}\,{\Lip{3}{y}}+{}\Biggl
({5}\,{\Lx^2}+{}\Biggl ({}-{2}\,{\Ly}+{38}\Biggr
){}\,{\Lx}-{76}\,{\Ly}+{2}\,{\Ly^2}-{13\over 3}\,{\pi^2}\Biggr
){}\,{\Lip{2}{x}} \nonumber \\
&&+{1\over 3}\,{\Lx^4}+{}\Biggl ({}-{2\over 3}\,{\Ly}-{13\over 9}\Biggr
){}\,{\Lx^3}+{}\Biggl ({269\over 18}+{6}\,{\pi^2}-{11\over
  2}\,{\LU}+{28}\,{\Ly}+{7\over 2}\,{\Ly^2}\Biggr ){}\,{\Lx^2} \nonumber \\
&&+{}\Biggl ({20\over 3}\,{\Ly^3}-{66}\,{\Ly^2}+{}\Biggl ({}-{269\over
  9}-{31\over 3}\,{\pi^2}+{11}\,{\LU}\Biggr ){}\,{\Ly}+{8\over
  9}\,{\pi^2}+{52}\,{\zeta_3}+{33\over 2}\,{\LU}-{31\over 2}\Biggr ){}\,{\Lx}
\nonumber \\
&&+{11\over 3}\,{\Ly^2}\,{\pi^2}+{}\Biggl ({}-{26}\,{\zeta_3}+{122\over
  3}\,{\pi^2}\Biggr ){}\,{\Ly}+{38}\,{\zeta_3}+{178\over
  45}\,{\pi^4}+{269\over 18}\,{\pi^2}-{11\over 2}\,{\pi^2}\,{\LU}\Biggr
]{}\,{\tmsost} \nonumber \\
&&+{}\Biggl [{8}\,{\Lip{4}{y}}+{8}\,{\Lip{4}{z}}+{}\Biggl
({}-{120}+{168}\,{\Ly}-{24}\,{\Lx}\Biggr
){}\,{\Lip{3}{x}}+{120}\,{\Lip{3}{y}}\,{\Lx} \nonumber \\
&&+{}\Biggl ({20}\,{\Lx^2}+{}\Biggl ({120}-{40}\,{\Ly}\Biggr
){}\,{\Lx}+{20}\,{\pi^2}\Biggr ){}\,{\Lip{2}{x}}-{8\over 3}\,{\Lx^4}+{}\Biggl
({40\over 3}\,{\Ly}-{184\over 9}\Biggr ){}\,{\Lx^3} \nonumber \\
&&+{}\Biggl ({472\over 9}+{14}\,{\Ly^2}+{122}\,{\Ly}-{22}\,{\LU}\Biggr
){}\,{\Lx^2}+{}\Biggl ({40\over 3}\,{\Ly^3}-{370\over 3}\,{\Ly^2}+{}\Biggl
({}-{76\over 3}\,{\pi^2}+{44}\,{\LU}-{320\over 9}\Biggr ){}\,{\Ly} \nonumber \\
&&-{112\over 9}\,{\pi^2}+{898\over 9}-{112}\,{\zeta_3}-{22}\,{\LU}\Biggr
){}\,{\Lx}-{20\over 3}\,{\Ly^4}+{740\over 9}\,{\Ly^3}+{}\Biggl ({320\over
  9}-{44}\,{\LU}\Biggr ){}\,{\Ly^2} \nonumber \\
&&+{}\Biggl ({44}\,{\LU}+{218\over 9}\,{\pi^2}+{104}\,{\zeta_3}-{1796\over
  9}\Biggr ){}\,{\Ly}+{96}\,{\zeta_3}+{104\over
  45}\,{\pi^4}+{4}-{22}\,{\pi^2}\,{\LU}+{60}\,{\pi^2}\Biggr ]{}\,{\one} \\
{E_{1;u}}&=&{}\Biggl [{11\over 6}\,{\Lx^3}+{}\Biggl ({3}\,{\LU}-{23\over
  6}-{6}\,{\Ly}\Biggr ){}\,{\Lx^2}+{}\Biggl ({7}\,{\Ly^2}+{}\Biggl
({}-{6}\,{\LU}+{20\over 3}\Biggr ){}\,{\Ly}+{11\over 6}\,{\pi^2}-{22\over
  3}+{3}\,{\LU}\Biggr ){}\,{\Lx} \nonumber \\
&&-{14\over 3}\,{\Ly^3}+{}\Biggl ({}-{20\over 3}+{6}\,{\LU}\Biggr
){}\,{\Ly^2}+{}\Biggl ({}-{6}\,{\LU}+{44\over 3}+{4\over 3}\,{\pi^2}\Biggr
){}\,{\Ly} \nonumber \\
&&-{2\over 9}\,{\zeta_3}+{8\over 3}\,{\pi^2}\,{\LU}-{121\over
  18}\,{\pi^2} +{3401\over 162}-{328\over 27}\,{\LU}\Biggr ]{}\,{\tpsost} \nonumber \\
&&+{}\Biggl [{11\over 18}\,{\Lx^3}+{}\Biggl ({}-{83\over
  18}+{\LU}-{2}\,{\Ly}\Biggr ){}\,{\Lx^2}+{}\Biggl ({2}\,{\Ly^2}+{}\Biggl
({83\over 9}-{2}\,{\LU}\Biggr ){}\,{\Ly}-{3}\,{\LU}+{5}+{11\over
  18}\,{\pi^2}\Biggr ){}\,{\Lx} \nonumber \\
&&-{2}\,{\pi^2}\,{\Ly}+{1\over 18}\,{\pi^2}\,{}\Biggl
({}-{83}+{18}\,{\LU}\Biggr ){}\Biggr ]{}\,{\tmsost} \nonumber \\
&&+{}\Biggl [{22\over 9}\,{\Lx^3}+{}\Biggl ({}-{46\over
  9}+{4}\,{\LU}-{8}\,{\Ly}\Biggr ){}\,{\Lx^2}+{}\Biggl ({28\over
  3}\,{\Ly^2}+{}\Biggl ({80\over 9}-{8}\,{\LU}\Biggr ){}\,{\Ly}+{22\over
  9}\,{\pi^2}-{76\over 9}+{4}\,{\LU}\Biggr ){}\,{\Lx} \nonumber \\
&&-{56\over 9}\,{\Ly^3}+{}\Biggl ({}-{80\over 9}+{8}\,{\LU}\Biggr
){}\,{\Ly^2}+{}\Biggl ({152\over 9}-{8}\,{\LU}+{16\over 9}\,{\pi^2}\Biggr
){}\,{\Ly}+{2}\,{\pi^2}\,{}\Biggl ({}-{5}+{2}\,{\LU}\Biggr ){}\Biggr
]{}\,{\one} \\
{E_{2;u}}&=&{}\Biggl [{4\over 3}\,{\Lip{3}{x}}-{4\over
  3}\,{\Lip{2}{x}}\,{\Lx}-{11\over 36}\,{\Lx^3}+{}\Biggl ({4\over
  3}\,{\Ly}-{13\over 18}-{1\over 2}\,{\LU}\Biggr ){}\,{\Lx^2} \nonumber \\
&&+{}\Biggl ({}-{7\over 2}\,{\Ly^2}+{}\Biggl ({1\over 3}+{3}\,{\LU}\Biggr
){}\,{\Ly}+{1\over 36}\,{\pi^2}-{31\over 6}\,{\LU}+{40\over 9}\Biggr
){}\,{\Lx} \nonumber \\
&&+{7\over 3}\,{\Ly^3}+{}\Biggl ({}-{1\over 3}-{3}\,{\LU}\Biggr
){}\,{\Ly^2}+{}\Biggl ({31\over 3}\,{\LU}-{5\over 9}\,{\pi^2}-{80\over
  9}\Biggr ){}\,{\Ly} \nonumber \\
&&-{25\over 9}\,{\zeta_3}-{13\over 9}\,{\pi^2}\,{\LU}-{11\over
  3}\,{\LU^2}+{206\over 27}\,{\LU}+{65\over 81}+{487\over 108}\,{\pi^2}\Biggr
]{}\,{\tpsost} \nonumber \\
&&+{}\Biggl [{}-{11\over 36}\,{\Lx^3}+{}\Biggl ({14\over 9}-{1\over
  2}\,{\LU}+{\Ly}\Biggr ){}\,{\Lx^2} +{}\Biggl ({}-{\Ly^2}+{}\Biggl
({}-{28\over 9}+{\LU}\Biggr ){}\,{\Ly}-{11\over 36}\,{\pi^2}-{1\over 3}\Biggr
){}\,{\Lx} \nonumber \\
&&+{\pi^2}\,{\Ly}-{1\over 18}\,{\pi^2}\,{}\Biggl
({}-{28}+{9}\,{\LU}\Biggr ){}\Biggr ]{}\,{\tmsost} \nonumber \\
&&-{\Lx^2}\,{\tpsou}+{2}\,{\Lx}\,{\tmsou} \nonumber \\
&&+{}\Biggl [{}-{11\over 9}\,{\Lx^3}+{}\Biggl ({14\over
  9}-{2}\,{\LU}+{4}\,{\Ly}\Biggr ){}\,{\Lx^2} \nonumber \\
&&+{}\Biggl ({}-{14\over 3}\,{\Ly^2}+{}\Biggl ({}-{40\over
  9}+{4}\,{\LU}\Biggr ){}\,{\Ly}+{38\over 9}-{2}\,{\LU}-{11\over
  9}\,{\pi^2}\Biggr ){}\,{\Lx}+{}\Biggl ({}-{76\over 9}+{4}\,{\LU}-{8\over
  9}\,{\pi^2}\Biggr ){}\,{\Ly} \nonumber \\
&&+{28\over 9}\,{\Ly^3}+{}\Biggl ({40\over 9}-{4}\,{\LU}\Biggr
){}\,{\Ly^2}-{\pi^2}\,{}\Biggl ({}-{5}+{2}\,{\LU}\Biggr ){}\Biggr ]{}\,{\one}
\\
{F_{1;u}}&=&{}\Biggl [{5\over 36}\,{\Lx^2}+{}\Biggl ({1\over
  3}\,{\LU}-{10\over 27}-{1\over 3}\,{\Ly}\Biggr ){}\,{\Lx}+{1\over
  3}\,{\Ly^2}+{}\Biggl ({20\over 27}-{2\over 3}\,{\LU}\Biggr ){}\,{\Ly}
\nonumber \\
&&-{20\over 27}\,{\LU}-{13\over 108}\,{\pi^2}+{1\over 3}\,{\LU^2}\Biggr
]{}\,{\tpsost} \\
{D_{2;u}}&=&{}\Biggl
[{}-{6}\,{\pi^2}-{6}\,{\Ly^2}+{12}\,{\Lx}\,{\Ly}-{6}\,{\Lx^2}\Biggr
]{}\,{\ttoss} -{6}\,{\Ly^2}\,{\ssott} \nonumber \\
&&+{6}\,{\Lx^2}\,{\tpsou}-{12}\,{\Lx}\,{\tmsou} \nonumber \\
&&+{}\Biggl [{}-{4}\,{\Lip{4}{y}}-{4}\,{\Lip{4}{z}}+{}\Biggl
({}-{4}-{4}\,{\Lx}+{36}\,{\Ly}\Biggr
){}\,{\Lip{3}{x}}+{28}\,{\Lip{3}{y}}\,{\Lx} \nonumber \\
&&+{}\Biggl ({6}\,{\Lx^2}-{12}\,{\Lx}\,{\Ly}+{6}\,{\pi^2}+{4}\,{\Lx}\Biggr
){}\,{\Lip{2}{x}}+{20}\,{\Ly^3}+{}\Biggl ({107\over
  3}+{3}\,{\Lx^2}-{22}\,{\LU}-{30}\,{\Lx}\Biggr ){}\,{\Ly^2} \nonumber \\
&&+{}\Biggl ({4}\,{\Lx^3}+{24}\,{\Lx^2}+{}\Biggl
({}-{4}\,{\pi^2}+{22}\,{\LU}-{107\over 3}\Biggr
){}\,{\Lx}-{57}+{36}\,{\zeta_3}+{22}\,{\LU}-{2\over 3}\,{\pi^2}\Biggr
){}\,{\Ly} \nonumber \\
&&-{\Lx^4}-{19\over 3}\,{\Lx^3}+{}\Biggl ({}-{11}\,{\LU}+{107\over 6}+{1\over
  3}\,{\pi^2}\Biggr ){}\,{\Lx^2}+{}\Biggl
({}-{11}\,{\LU}-{32}\,{\zeta_3}+{57\over 2}-{20\over 3}\,{\pi^2}\Biggr
){}\,{\Lx} \nonumber \\
&&-{43417\over 324}-{43\over
  6}\,{\pi^2}\,{\LU}-{52}\,{\zeta_3}\,{\LU}+{515\over 9}\,{\zeta_3}+{251\over
  9}\,{\pi^2}+{1141\over 27}\,{\LU}+{65\over 36}\,{\pi^4}\Biggr
]{}\,{\tpsost} \nonumber \\
&&+{}\Biggl
[{}-{20}\,{\Lip{4}{y}}-{48}\,{\Lip{4}{x}}+{20}\,{\Lip{4}{z}}+{}\Biggl
({12}\,{\Ly}-{16}+{12}\,{\Lx}\Biggr ){}\,{\Lip{3}{x}} \nonumber \\
&&+{}\Biggl
({4}\,{\Ly^2}+{}\Biggl ({}-{4}\,{\Lx}-{32}\Biggr
){}\,{\Ly}+{2}\,{\Lx^2}+{16}\,{\Lx}-{10\over 3}\,{\pi^2}\Biggr
){}\,{\Lip{2}{x}} \nonumber \\
&&+{}\Biggl ({}-{12}\,{\Lx}+{24}\,{\Ly}-{32}\Biggr
){}\,{\Lip{3}{y}}+{4}\,{\Ly^3}\,{\Lx}+{}\Biggl ({}-{94\over
  3}\,{\Lx}+{\Lx^2}+{4\over 3}\,{\pi^2}\Biggr ){}\,{\Ly^2} \nonumber \\
&&+{}\Biggl ({46\over 3}\,{\Lx^2}+{}\Biggl ({}-{20\over 3}\,{\pi^2}+{22\over
  3}\,{\LU}-{251\over 9}\Biggr ){}\,{\Lx}-{12}\,{\zeta_3}+{62\over
  3}\,{\pi^2}\Biggr ){}\,{\Ly} \nonumber \\
&&-{13\over 9}\,{\Lx^3}+{}\Biggl ({251\over 18}+{3}\,{\pi^2}-{11\over
  3}\,{\LU}\Biggr ){}\,{\Lx^2}+{}\Biggl ({}-{16\over
  9}\,{\pi^2}+{24}\,{\zeta_3}-{57\over 2}+{11}\,{\LU}\Biggr ){}\,{\Lx} \nonumber \\
&&+{16}\,{\zeta_3}+{251\over 18}\,{\pi^2}-{11\over
  3}\,{\pi^2}\,{\LU}+{94\over 45}\,{\pi^4}\Biggr ]{}\,{\tmsost} \nonumber \\
&&+{}\Biggl [{}-{16}\,{\Lip{4}{y}}-{16}\,{\Lip{4}{z}}+{}\Biggl
({}-{64}+{48}\,{\Ly}\Biggr ){}\,{\Lip{3}{x}}+{48}\,{\Lip{3}{y}}\,{\Lx} \nonumber \\
&&+{}\Biggl
({}-{16}\,{\Lx}\,{\Ly}+{64}\,{\Lx}+{8}\,{\Lx^2}+{8}\,{\pi^2}\Biggr
){}\,{\Lip{2}{x}}+{272\over 9}\,{\Ly^3} \nonumber \\
&&+{}\Biggl ({4}\,{\Lx^2}+{344\over 9}-{136\over 3}\,{\Lx}-{88\over
  3}\,{\LU}\Biggr ){}\,{\Ly^2}+{}\Biggl ({8}\,{\Lx^3}+{184\over
  3}\,{\Lx^2}+{}\Biggl ({88\over 3}\,{\LU}-{344\over 9}-{16\over
  3}\,{\pi^2}\Biggr ){}\,{\Lx} \nonumber \\
&&+{88\over 3}\,{\LU}+{32\over 9}\,{\pi^2}+{48}\,{\zeta_3}-{1076\over
  9}\Biggr ){}\,{\Ly}-{2}\,{\Lx^4}-{112\over 9}\,{\Lx^3}+{}\Biggl
({}-{44\over 3}\,{\LU}+{298\over 9}\Biggr ){}\,{\Lx^2} \nonumber \\
&&+{}\Biggl ({}-{76\over 9}\,{\pi^2}-{44\over 3}\,{\LU}+{538\over
  9}-{48}\,{\zeta_3}\Biggr ){}\,{\Lx}+{48}\,{\pi^2}+{48}\,{\zeta_3}-{44\over
  3}\,{\pi^2}\,{\LU}+{92\over 45}\,{\pi^4}\Biggr ]{}\,{\one} \\
{E_{3;u}}&=&{}\Biggl [{}-{8\over 3}\,{\Ly^3}+{}\Biggl ({4}\,{\LU}-{26\over
  3}+{4}\,{\Lx}\Biggr ){}\,{\Ly^2}+{}\Biggl ({}-{4}\,{\Lx^2}+{}\Biggl
({26\over 3}-{4}\,{\LU}\Biggr ){}\,{\Lx}+{4\over
  3}\,{\pi^2}+{10}-{4}\,{\LU}\Biggr ){}\,{\Ly} \nonumber \\
&&+{4\over 3}\,{\Lx^3}+{}\Biggl ({2}\,{\LU}-{13\over 3}\Biggr
){}\,{\Lx^2}+{}\Biggl ({4\over 3}\,{\pi^2}+{2}\,{\LU}-{5}\Biggr
){}\,{\Lx} \nonumber \\
&&-{2\over 9}\,{\zeta_3}+{5\over 3}\,{\pi^2}\,{\LU}+{3401\over
  162}-{56\over 9}\,{\pi^2}-{202\over 27}\,{\LU}\Biggr ]{}\,{\tpsost} \nonumber \\
&&+{}\Biggl [{4\over 3}\,{\Ly^2}\,{\Lx}+{}\Biggl ({}-{4\over
  3}\,{\Lx^2}+{}\Biggl ({}-{4\over 3}\,{\LU}+{74\over 9}\Biggr
){}\,{\Lx}-{4\over 3}\,{\pi^2}\Biggr ){}\,{\Ly}+{4\over 9}\,{\Lx^3}+{}\Biggl
({}-{37\over 9}+{2\over 3}\,{\LU}\Biggr ){}\,{\Lx^2} \nonumber \\
&&+{}\Biggl ({}-{2}\,{\LU}+{5}+{4\over 9}\,{\pi^2}\Biggr ){}\,{\Lx}+{1\over
  9}\,{\pi^2}\,{}\Biggl ({}-{37}+{6}\,{\LU}\Biggr ){}\Biggr ]{}\,{\tmsost} \nonumber \\
&&+{}\Biggl [{}-{32\over 9}\,{\Ly^3}+{}\Biggl ({16\over 3}\,{\LU}+{16\over
  3}\,{\Lx}-{104\over 9}\Biggr ){}\,{\Ly^2}+{}\Biggl ({}-{16\over
  3}\,{\Lx^2}+{}\Biggl ({}-{16\over 3}\,{\LU}+{104\over 9}\Biggr ){}\,{\Lx}
\nonumber \\
&&+{152\over 9}-{16\over 3}\,{\LU}+{16\over 9}\,{\pi^2}\Biggr
){}\,{\Ly}+{16\over 9}\,{\Lx^3}+{}\Biggl ({8\over 3}\,{\LU}-{52\over 9}\Biggr
){}\,{\Lx^2}+{}\Biggl ({8\over 3}\,{\LU}-{76\over 9}+{16\over
  9}\,{\pi^2}\Biggr ){}\,{\Lx} \nonumber \\
&&+{4\over 3}\,{\pi^2}\,{}\Biggl ({2}\,{\LU}-{7}\Biggr ){}\Biggr
]{}\,{\one} \\
{E_{4;u}}&=&{}\Biggl [{8\over 3}\,{\Lx^3}+{}\Biggl ({8}\,{\LU}-{26\over
  3}-{8}\,{\Ly}\Biggr ){}\,{\Lx^2}+{}\Biggl ({8}\,{\Ly^2}+{}\Biggl ({52\over
  3}-{16}\,{\LU}\Biggr ){}\,{\Ly}+{8}\,{\LU}+{8\over 3}\,{\pi^2}-{10}\Biggr
){}\,{\Lx} \nonumber \\
&&-{16\over 3}\,{\Ly^3}+{}\Biggl ({}-{52\over 3}+{16}\,{\LU}\Biggr
){}\,{\Ly^2}+{}\Biggl ({8\over 3}\,{\pi^2}-{16}\,{\LU}+{20}\Biggr
){}\,{\Ly} \nonumber \\
&&-{112\over 9}\,{\pi^2}-{908\over 27}\,{\LU}-{4\over
  9}\,{\zeta_3}+{22\over 3}\,{\pi^2}\,{\LU}+{3401\over 81}\Biggr
]{}\,{\tpsost} \nonumber \\
&&+{}\Biggl [{8\over 9}\,{\Lx^3}+{}\Biggl ({8\over 3}\,{\LU}-{74\over
  9}-{8\over 3}\,{\Ly}\Biggr ){}\,{\Lx^2} \nonumber \\
&&+{}\Biggl ({8\over
  3}\,{\Ly^2}+{}\Biggl ({}-{16\over 3}\,{\LU}+{148\over 9}\Biggr
){}\,{\Ly}+{8\over 9}\,{\pi^2}+{10}-{8}\,{\LU}\Biggr ){}\,{\Lx} \nonumber \\
&&-{8\over 3}\,{\pi^2}\,{\Ly}+{2\over 9}\,{\pi^2}\,{}\Biggl
({12}\,{\LU}-{37}\Biggr ){}\Biggr ]{}\,{\tmsost} \nonumber \\
&&+{}\Biggl [{32\over 9}\,{\Lx^3}+{}\Biggl ({}-{104\over 9}-{32\over
  3}\,{\Ly}+{32\over 3}\,{\LU}\Biggr ){}\,{\Lx^2} \nonumber \\
&&+{}\Biggl ({32\over 3}\,{\Ly^2}+{}\Biggl ({}-{64\over 3}\,{\LU}+{208\over
  9}\Biggr ){}\,{\Ly}-{152\over 9}+{32\over 3}\,{\LU}+{32\over
  9}\,{\pi^2}\Biggr ){}\,{\Lx} \nonumber \\
&&-{64\over 9}\,{\Ly^3}+{}\Biggl ({}-{208\over 9}+{64\over 3}\,{\LU}\Biggr
){}\,{\Ly^2}+{}\Biggl ({}-{64\over 3}\,{\LU}+{304\over 9}+{32\over
  9}\,{\pi^2}\Biggr ){}\,{\Ly} \nonumber \\
&&+{8\over 3}\,{\pi^2}\,{}\Biggl ({}-{7}+{4}\,{\LU}\Biggr ){}\Biggr ]{}\,{\one}\\
{F_{2;u}}&=&{}\Biggl [{4\over 27}\,{\pi^2}+{32\over 9}\,{\LU^2}-{160\over
  27}\,{\LU}\Biggr ]{}\,{\tpsost}
\end{eqnarray}

\section{Finite one-loop contributions}
\subsection{$s$-channel}
\label{app:sone}

\begin{eqnarray}
{G_{1;s}}&=&{}\Biggl
[{14}\,{\Lx^4}+{28}\,{\Lx^3}+{8}\,{\Lx^2}\,{\Ly^2}+{56}\,{\Lx^2}\,{\pi^2}-{48}\,{\Lx^2}+{12}\,{\Lx^2}\,{\Ly}+{32}\,{\Lx}\,{\Ly}\,{\pi^2}+{80}\,{\pi^2}\,{\Lx}
\nonumber \\
&&+{2}\,{\Ly^4}+{12}\,{\Ly^3}-{10}\,{\Ly^2}+{8}\,{\Ly^2}\,{\pi^2}+{26}\,{\pi^2}+{24}\,{\pi^2}\,{\Ly}-{84}\,{\Ly}+{102}\Biggr
]{}\,{\tou} \nonumber \\
&&+{8}\,{\Lx}\,{}\Biggl
[{\Lx^3}+{\Lx^2}+{4}\,{\pi^2}\,{\Lx}+{2}\,{\pi^2}\Biggr
]{}\,{\ttouu}+{2}\,{\Lx^2}\,{}\Biggl [{\Lx^2}+{4}\,{\pi^2}\Biggr
]{}\,{\tttouuu} \nonumber \\
&&+{}\Biggl
[{32}\,{\Lx^3}+{8}\,{\Lx^2}\,{\Ly^2}+{80}\,{\pi^2}\,{\Lx}+{32}\,{\Lx}\,{\Ly}\,{\pi^2}+{8}\,{\Ly^2}\,{\Lx}
\nonumber \\
&&+{8}\,{\Ly^4}+{32}\,{\Ly^2}\,{\pi^2}-{32}\,{\Ly^2}-{4}-{56}\,{\Ly}+{24}\,{\pi^2}\Biggr
] + \Biggl\{t \leftrightarrow u \Biggr\} \\
{G_{2;s}}&=&{}\Biggl [{}-{10}\,{\Lx^4}-{106\over
  3}\,{\Lx^3}-{8}\,{\Lx^3}\,{\Ly}-{2}\,{\Lx^2}\,{\Ly^2}-{52}\,{\Lx^2}\,{\pi^2}-{44\over 3}\,{\Lx^2}\,{\Ls}+{6}\,{\Lx^2}-{40\over 3}\,{\Lx^2}\,{\Ly} \nonumber \\
&&-{4}\,{\Ly^3}\,{\Lx}-{56\over
  3}\,{\Ly^2}\,{\Lx}-{32}\,{\Lx}\,{\Ly}\,{\pi^2}+{8}\,{\Lx}\,{\Ly}-{80}\,{\pi^2}\,{\Lx}+{140\over 3}\,{\Lx}-{20\over 3}\,{\Ly^3}-{22\over 3}\,{\Ly^2}\,{\Ls} \nonumber \\
&&-{20}\,{\Ly^2}-{6}\,{\Ly^2}\,{\pi^2}-{18}\,{\pi^2}\,{\Ly}-{22}\,{\Ly}\,{\Ls}+{140\over 3}\,{\Ly}-{4}+{154\over 3}\,{\Ls}-{40}\,{\pi^2}\Biggr ]{}\,{\tou} \nonumber \\
&&-{8}\,{\Lx}\,{}\Biggl
[{\Lx^3}+{\Lx^2}+{4}\,{\pi^2}\,{\Lx}+{2}\,{\pi^2}\Biggr
]{}\,{\ttouu}-{2}\,{\Lx^2}\,{}\Biggl [{\Lx^2}+{4}\,{\pi^2}\Biggr
]{}\,{\tttouuu} \nonumber \\
&&+{}\Biggl
[{}-{8}\,{\Lx^3}\,{\Ly}-{32}\,{\Lx^3}+{4}\,{\Lx^2}\,{\pi^2}-{4}\,{\Lx^2}\,{\Ly^2}-{52\over
  3}\,{\Lx^2}\,{\Ly}-{8}\,{\Ly^2}\,{\Lx}-{40\over 3}\,{\Lx}\,{\Ly} \nonumber \\
&&-{32}\,{\Lx}\,{\Ly}\,{\pi^2}-{76}\,{\pi^2}\,{\Lx}-{44\over
  3}\,{\Lx}\,{\Ls}-{4}\,{\Ly^4}+{8\over 3}\,{\Ly^3}+{8\over
  3}\,{\Ly^2}-{44\over 3}\,{\Ly^2}\,{\Ls}-{32}\,{\Ly^2}\,{\pi^2} \nonumber \\
&&+{28}\,{\Ly}+{4}-{24}\,{\pi^2}\Biggr ] + \Biggl\{t \leftrightarrow u \Biggr\} \\
{G_{3;s}}&=&{}\Biggl [{2}\,{\Lx^4}+{2}\,{\Lx^3}\,{\Ly}+{22\over
  3}\,{\Lx^3}+{11\over
  3}\,{\Lx^2}\,{\Ls}+{2}\,{\Lx^2}\,{\Ly^2}+{10}\,{\Lx^2}\,{\Ly}+{68\over
  9}\,{\Lx^2}+{13}\,{\Lx^2}\,{\pi^2} \nonumber \\
&&+{20\over 3}\,{\Ly^2}\,{\Lx}+{6}\,{\Lx}\,{\Ly}\,{\pi^2}+{100\over
  9}\,{\Lx}\,{\Ly}+{22\over 3}\,{\Lx}\,{\Ly}\,{\Ls}+{110\over
  9}\,{\Lx}\,{\Ls}+{50\over 3}\,{\pi^2}\,{\Lx}+{50\over
  9}\,{\Ly^2} \nonumber \\
&&+{2}\,{\Ly^2}\,{\pi^2}+{8\over 3}\,{\pi^2}\,{\Ly}+{110\over
  9}\,{\Ly}\,{\Ls}+{2}+{121\over 18}\,{\Ls^2}-{11\over
  3}\,{\pi^2}\,{\Ls}+{1\over 2}\,{\pi^4}+{13\over 2}\,{\pi^2}\Biggr
]{}\,{\tou} \nonumber \\
&&+{2}\,{\Lx}\,{}\Biggl
[{\Lx^3}+{\Lx^2}+{4}\,{\pi^2}\,{\Lx}+{2}\,{\pi^2}\Biggr ]{}\,{\ttouu}+{1\over
  2}\,{\Lx^2}\,{}\Biggl [{\Lx^2}+{4}\,{\pi^2}\Biggr ]{}\,{\tttouuu} \nonumber \\
&&+{}\Biggl [{4}\,{\Lx^3}\,{\Ly}+{8}\,{\Lx^3}-{2}\,{\Lx^2}\,{\pi^2}+{26\over
  3}\,{\Lx^2}\,{\Ly}+{2}\,{\Ly^2}\,{\Lx}+{8}\,{\Lx}\,{\Ly}\,{\pi^2}+{20\over
  3}\,{\Lx}\,{\Ly}+{22\over 3}\,{\Lx}\,{\Ls} \nonumber \\
&&+{18}\,{\pi^2}\,{\Lx}-{4\over 3}\,{\Ly^3}+{20\over 3}\,{\Ly^2}+{22\over
  3}\,{\Ly^2}\,{\Ls}+{8}\,{\Ly^2}\,{\pi^2}+{6}\,{\pi^2}\Biggr ] + \Biggl\{t \leftrightarrow u \Biggr\} \\
{X_{1;s}}&=&{2\over 3}\,{}\Biggl ({3}\,{\Ly}+{\Ly^2}-{7}+{2}\,{\Lx^2}\Biggr
){}\,{}\Biggl ({2}\,{\Ls}+{\Ly}+{\Lx}\Biggr ){}\,{\tou} \nonumber \\
&&+{}\Biggl ({4\over 3}\,{\Lx^2}\,{\Ly}+{8\over 3}\,{\Lx}\,{\Ls}+{4\over
  3}\,{\Lx}\,{\Ly}+{4\over 3}\,{\Ly^3}+{4\over 3}\,{\Ly^2}+{8\over
  3}\,{\Ly^2}\,{\Ls}\Biggr ) + \Biggl\{t \leftrightarrow u \Biggr\} \\
{X_{2;s}}&=&{}-{1\over 9}\,{}\Biggl ({2}\,{\Ls}+{\Ly}+{\Lx}\Biggr
){}\,{}\Biggl
({11}\,{\Ls}+{3}\,{\Lx^2}+{6}\,{\Lx}\,{\Ly}+{10}\,{\Ly}+{10}\,{\Lx}-{3}\,{\pi^2}\Biggr
){}\,{\tou} \nonumber \\
&&+{}\Biggl ({}-{2\over 3}\,{\Lx}\,{\Ly}-{2\over 3}\,{\Ly^2}-{2\over
  3}\,{\Lx^2}\,{\Ly}-{4\over 3}\,{\Ly^2}\,{\Ls}-{4\over
  3}\,{\Lx}\,{\Ls}-{2\over 3}\,{\Ly^3}\Biggr ) + \Biggl\{t \leftrightarrow u \Biggr\} \\
{X_{3;s}}&=&{1\over 18}\,{}\Biggl ({2}\,{\Ls}+{\Ly}
+{\Lx}\Biggr ){^2}\,{\tou} + \Biggl\{t \leftrightarrow u \Biggr\} \\
{X_{4;s}}&=&{ }\Biggl ({}-{32\over 3}\,{\Ly}\,{\pi^2}+{32\over
  3}\,{\Ls}\,{\Lx^2}+{16}\,{\Ls}\,{\Ly}-{64\over
  3}\,{\Lx}\,{\pi^2}-{16}\,{\pi^2}+{16\over 3}\,{\Ls}\,{\Ly^2}-{112\over
  3}\,{\Ls}\Biggr ){}\,{\tou} \nonumber \\
&&+{}\Biggl ({32\over 3}\,{\Ls}\,{\Lx^2}+{32\over 3}\,{\Ls}\,{\Ly}-{64\over
  3}\,{\Lx}\,{\pi^2}-{32\over 3}\,{\pi^2}\Biggr ) + \Biggl\{t \leftrightarrow u \Biggr\} \\
{X_{5;s}}&=&{ }\Biggl ({32\over 9}\,{\Ls^2}+{32\over 9}\,{\pi^2}\Biggr
){}\,{\tou} + \Biggl\{t \leftrightarrow u \Biggr\}
\end{eqnarray}

\subsection{$u$-channel}
\label{app:uone}

\begin{eqnarray}
{G_{1;u}}&=&{}\Biggl [{8}\,{\Lx^4}+{}\Biggl ({}-{32}\,{\Ly}+{8}\Biggr
){}\,{\Lx^3}+{}\Biggl ({48}\,{\Ly^2}-{24}\,{\Ly}+{16}\,{\pi^2}\Biggr
){}\,{\Lx^2} \nonumber \\
&&+{}\Biggl
({}-{32}\,{\Ly^3}+{24}\,{\Ly^2}-{32}\,{\pi^2}\,{\Ly}+{8}\,{\pi^2}\Biggr
){}\,{\Lx} \nonumber \\
&&+{8}\,{\Ly^4}-{8}\,{\Ly^3}+{16}\,{\Ly^2}\,{\pi^2}-{8}\,{\pi^2}\,{\Ly}+{8}\,{\pi^4}\Biggr
]{}\,{\ttoss} \nonumber \\
&&+{}\Biggl[{8}\,{}\,{\Ly^4}-{8}\,{\Ly^3}
+{32}\,{\pi^2}\,{\Ly^2}-{16}\,{\pi^2}\,{\Ly}\Biggr ]\,{\ssott}\nonumber \\
&&+{}\Biggl [{2}\,{\Lx^4}-{8}\,{\Lx^3}\,{\Ly}+{}\Biggl
({4}\,{\pi^2}+{12}\,{\Ly^2}\Biggr ){}\,{\Lx^2}+{}\Biggl
({}-{8}\,{\Ly^3}-{8}\,{\pi^2}\,{\Ly}\Biggr
){}\,{\Lx} \nonumber \\
&&+{2}\,{\Ly^4}+{4}\,{\Ly^2}\,{\pi^2}+{2}\,{\pi^4}\Biggr ]{}\,{\tttosss}
+{}\Biggl[{2}\,{\Ly^4} + {8}\,{\pi^2}\,{\Ly^2}\Biggr] {}\,{\sssottt} \nonumber \\
&&+{}\Biggl [{8}\,{\Lx^4}+{}\Biggl ({}-{32}\,{\Ly}+{20}\Biggr
){}\,{\Lx^3}+{}\Biggl ({16}\,{\pi^2}+{56}\,{\Ly^2}-{66}\,{\Ly}-{29}\Biggr
){}\,{\Lx^2} \nonumber \\
&&+{}\Biggl ({}-{48}\,{\Ly^3}+{78}\,{\Ly^2}+{}\Biggl
({58}-{32}\,{\pi^2}\Biggr ){}\,{\Ly}-{42}+{20}\,{\pi^2}\Biggr ){}\,{\Lx} \nonumber \\
&&+{24}\,{\Ly^4}-{52}\,{\Ly^3}+{}\Biggl ({56}\,{\pi^2}-{58}\Biggr
){}\,{\Ly^2}+{}\Biggl ({}-{66}\,{\pi^2}+{84}\Biggr
){}\,{\Ly} \nonumber \\
&&+{102}+{8}\,{\pi^4}-{29}\,{\pi^2}\Biggr ]{}\,{\tpsost} \nonumber \\
&&+{}\Biggl [{6}\,{\Lx^4}+{}\Biggl ({}-{24}\,{\Ly}+{8}\Biggr
){}\,{\Lx^3}+{}\Biggl ({36}\,{\Ly^2}-{19}-{30}\,{\Ly}+{12}\,{\pi^2}\Biggr
){}\,{\Lx^2} \nonumber \\
&&+{}\Biggl ({}-{24}\,{\Ly^3}+{30}\,{\Ly^2}+{}\Biggl
({38}-{24}\,{\pi^2}\Biggr ){}\,{\Ly}+{8}\,{\pi^2}+{42}\Biggr
){}\,{\Lx} \nonumber \\
&&-{12}\,{\Ly^2}\,{\pi^2}+{2}\,{\pi^2}\,{\Ly}+{3}\,{\pi^2}\,{}\Biggl
({2}\,{\pi^2}-{3}\Biggr ){}\Biggr ]{}\,{\tmsost} \nonumber \\
&&+{}\Biggl [{8}\,{\Lx^4}+{}\Biggl ({32}-{32}\,{\Ly}\Biggr
){}\,{\Lx^3}+{}\Biggl ({}-{104}\,{\Ly}+{64}\,{\Ly^2}-{32}+{16}\,{\pi^2}\Biggr
){}\,{\Lx^2} \nonumber \\
&&+{}\Biggl ({}-{64}\,{\Ly^3}+{120}\,{\Ly^2}+{}\Biggl
({}-{32}\,{\pi^2}+{64}\Biggr ){}\,{\Ly}-{56}+{32}\,{\pi^2}\Biggr ){}\,{\Lx}
\nonumber \\
&&+{32}\,{\Ly^4}-{80}\,{\Ly^3}+{64}\,{}\Biggl ({\pi}-{1}\Biggr ){}\,{}\Biggl
({\pi}+{1}\Biggr ){}\,{\Ly^2} \nonumber \\
&&+{}\Biggl ({}-{104}\,{\pi^2}+{112}\Biggr
){}\,{\Ly}-{8}+{8}\,{\pi^4}-{32}\,{\pi^2}\Biggr ]{}\,{\one}  \\
{G_{2;u}}&=&{}\Biggl [{}-{8}\,{\Lx^4}+{}\Biggl ({32}\,{\Ly}-{8}\Biggr
){}\,{\Lx^3}+{}\Biggl ({}-{48}\,{\Ly^2}+{24}\,{\Ly}-{16}\,{\pi^2}\Biggr
){}\,{\Lx^2} \nonumber \\
&&+{}\Biggl
({32}\,{\Ly^3}-{24}\,{\Ly^2}+{32}\,{\pi^2}\,{\Ly}-{8}\,{\pi^2}\Biggr
){}\,{\Lx} \nonumber \\
&&-{8}\,{\Ly^4}+{8}\,{\Ly^3}-{16}\,{\Ly^2}\,{\pi^2}+{8}\,{\pi^2}\,{\Ly}-{8}\,{\pi^4}\Biggr
]{}\,{\ttoss} \nonumber \\
&&-{}\Biggl[{8}\,{}\,{\Ly^4}-{8}\,{\Ly^3}
+{32}\,{\pi^2}\,{\Ly^2}-{16}\,{\pi^2}\,{\Ly}\Biggr ]\,{\ssott}\nonumber \\
&&+{}\Biggl [{}-{2}\,{\Lx^4}+{8}\,{\Lx^3}\,{\Ly}+{}\Biggl
({}-{4}\,{\pi^2}-{12}\,{\Ly^2}\Biggr ){}\,{\Lx^2}+{}\Biggl
({8}\,{\Ly^3}+{8}\,{\pi^2}\,{\Ly}\Biggr ){}\,{\Lx} \nonumber \\
&&-{2}\,{\Ly^4}-{4}\,{\Ly^2}\,{\pi^2}-{2}\,{\pi^4}\Biggr ]{}\,{\tttosss} 
-{}\Biggl[{2}\,{\Ly^4} + {8}\,{\pi^2}\,{\Ly^2}\Biggr] {}\,{\sssottt}\nonumber \\
&&+{}\Biggl [{}-{5}\,{\Lx^4}+{}\Biggl ({}-{21}+{26}\,{\Ly}\Biggr
){}\,{\Lx^3}+{}\Biggl
({}-{11}\,{\LU}-{7}-{50}\,{\Ly^2}-{13}\,{\pi^2}+{79}\,{\Ly}\Biggr
){}\,{\Lx^2} \nonumber \\
&&+{}\Biggl ({48}\,{\Ly^3}-{111}\,{\Ly^2}+{}\Biggl
({6}+{44}\,{\pi^2}+{22}\,{\LU}\Biggr ){}\,{\Ly}+{140\over
  3}-{30}\,{\pi^2}-{11}\,{\LU}\Biggr ){}\,{\Lx} \nonumber \\
&&-{24}\,{\Ly^4}+{74}\,{\Ly^3}+{}\Biggl
({}-{6}-{56}\,{\pi^2}-{22}\,{\LU}\Biggr ){}\,{\Ly^2}+{}\Biggl ({}-{280\over
  3}+{85}\,{\pi^2}+{22}\,{\LU}\Biggr ){}\,{\Ly} \nonumber \\
&&+{154\over
  3}\,{\LU}-{4}+{7}\,{\pi^2}-{11}\,{\pi^2}\,{\LU}-{8}\,{\pi^4}\Biggr
]{}\,{\tpsost} \nonumber \\
&&+{}\Biggl [{}-{5}\,{\Lx^4}+{}\Biggl ({22}\,{\Ly}-{43\over 3}\Biggr
){}\,{\Lx^3}+{}\Biggl ({121\over
  3}\,{\Ly}-{11}\,{\pi^2}-{36}\,{\Ly^2}+{13}-{11\over 3}\,{\LU}\Biggr
){}\,{\Lx^2} \nonumber \\
&&+{}\Biggl ({24}\,{\Ly^3}-{121\over 3}\,{\Ly^2}+{}\Biggl ({}-{26}+{22\over
  3}\,{\LU}+{20}\,{\pi^2}\Biggr ){}\,{\Ly}-{52\over
  3}\,{\pi^2}+{11}\,{\LU}\Biggr
){}\,{\Lx} \nonumber \\
&&+{12}\,{\Ly^2}\,{\pi^2}-{5}\,{\pi^2}\,{\Ly}-{1\over
  3}\,{\pi^2}\,{}\Biggl ({18}\,{\pi^2}+{11}\,{\LU}-{3}\Biggr ){}\Biggr
]{}\,{\tmsost} \nonumber \\
&&+{}\Biggl [{}-{4}\,{\Lx^4}+{}\Biggl ({}-{88\over 3}+{24}\,{\Ly}\Biggr
){}\,{\Lx^3}+{}\Biggl ({}-{44\over 3}\,{\LU}+{8\over
  3}-{56}\,{\Ly^2}-{12}\,{\pi^2}+{340\over 3}\,{\Ly}\Biggr ){}\,{\Lx^2} \nonumber \\
&&+{}\Biggl ({64}\,{\Ly^3}-{164}\,{\Ly^2}+{}\Biggl ({64\over
  3}+{48}\,{\pi^2}+{88\over 3}\,{\LU}\Biggr ){}\,{\Ly}+{28}-{124\over
  3}\,{\pi^2}-{44\over 3}\,{\LU}\Biggr ){}\,{\Lx} \nonumber \\
&&-{32}\,{\Ly^4}+{328\over 3}\,{\Ly^3}+{}\Biggl ({}-{64\over
  3}-{64}\,{\pi^2}-{88\over 3}\,{\LU}\Biggr ){}\,{\Ly^2}+{}\Biggl
({}-{56}+{364\over 3}\,{\pi^2}+{88\over 3}\,{\LU}\Biggr ){}\,{\Ly} \nonumber \\
&&+{8}+{8\over 3}\,{\pi^2}-{44\over 3}\,{\pi^2}\,{\LU}-{8}\,{\pi^4}\Biggr
]{}\,{\one}  \\
{G_{3;u}}&=&{}\Biggl [{2}\,{\Lx^4}+{}\Biggl ({}-{8}\,{\Ly}+{2}\Biggr
){}\,{\Lx^3}+{}\Biggl ({12}\,{\Ly^2}-{6}\,{\Ly}+{4}\,{\pi^2}\Biggr
){}\,{\Lx^2} \nonumber \\
&&+{}\Biggl
({}-{8}\,{\Ly^3}+{6}\,{\Ly^2}-{8}\,{\pi^2}\,{\Ly}+{2}\,{\pi^2}\Biggr
){}\,{\Lx} \nonumber \\
&&+{2}\,{\Ly^4}-{2}\,{\Ly^3}+{4}\,{\Ly^2}\,{\pi^2}-{2}\,{\pi^2}\,{\Ly}+{2}\,{\pi^4}\Biggr
]{}\,{\ttoss} \nonumber \\
&&+{}\Biggl[{2}\,{}\,{\Ly^4}-{2}\,{\Ly^3}
+{8}\,{\pi^2}\,{\Ly^2}-{4}\,{\pi^2}\,{\Ly}\Biggr ]\,{\ssott}\nonumber \nonumber \\
&&+{}\Biggl [{1\over 2}\,{\Lx^4}-{2}\,{\Lx^3}\,{\Ly}+{}\Biggl
({\pi^2}+{3}\,{\Ly^2}\Biggr ){}\,{\Lx^2}+{}\Biggl
({}-{2}\,{\Ly^3}-{2}\,{\pi^2}\,{\Ly}\Biggr ){}\,{\Lx} \nonumber \\
&&+{1\over 2}\,{\Ly^4}+{\Ly^2}\,{\pi^2}+{1\over 2}\,{\pi^4}\Biggr
]{}\,{\tttosss} +{}\Biggl[{1\over 2}\,{\Ly^4} + {2}\,{\pi^2}\,{\Ly^2}\Biggr]
{}\,{\sssottt} \nonumber \\
&&+{}\Biggl [{\Lx^4}+{}\Biggl ({11\over 3}-{5}\,{\Ly}\Biggr
){}\,{\Lx^3}+{}\Biggl ({11\over 6}\,{\LU}+{59\over 9}+{11}\,{\Ly^2}+{9\over
  2}\,{\pi^2}-{58\over 3}\,{\Ly}\Biggr ){}\,{\Lx^2} \nonumber \\
&&+{}\Biggl ({}-{12}\,{\Ly^3}+{36}\,{\Ly^2}+{}\Biggl
({}-{14}\,{\pi^2}-{218\over 9}-{11}\,{\LU}\Biggr ){}\,{\Ly}+{110\over
  9}\,{\LU}+{41\over 3}\,{\pi^2}\Biggr ){}\,{\Lx} \nonumber \nonumber \\
&&+{6}\,{\Ly^4}-{24}\,{\Ly^3}+{}\Biggl ({11}\,{\LU}+{218\over
  9}+{14}\,{\pi^2}\Biggr ){}\,{\Ly^2}+{}\Biggl ({}-{26}\,{\pi^2}-{220\over
  9}\,{\LU}\Biggr ){}\,{\Ly} \nonumber \\
&&+{121\over 18}\,{\LU^2}+{2}\,{\pi^4}+{11\over 2}\,{\pi^2}\,{\LU}+{59\over
  9}\,{\pi^2}+{2}\Biggr ]{}\,{\tpsost} \nonumber \\
&&+{}\Biggl [{\Lx^4}+{}\Biggl ({11\over 3}-{5}\,{\Ly}\Biggr
){}\,{\Lx^3}+{}\Biggl ({1}-{38\over 3}\,{\Ly}+{5\over
  2}\,{\pi^2}+{9}\,{\Ly^2}+{11\over 6}\,{\LU}\Biggr ){}\,{\Lx^2} \nonumber \\
&&+{}\Biggl ({}-{6}\,{\Ly^3}+{38\over 3}\,{\Ly^2}+{}\Biggl
({}-{4}\,{\pi^2}-{11\over 3}\,{\LU}-{2}\Biggr ){}\,{\Ly}+{11\over
  3}\,{\pi^2}\Biggr
){}\,{\Lx} \nonumber \\
&&-{3}\,{\Ly^2}\,{\pi^2}+{2}\,{\pi^2}\,{\Ly}+{1\over
  6}\,{\pi^2}\,{}\Biggl ({9}\,{\pi^2}+{11}\,{\LU}-{6}\Biggr ){}\Biggr
]{}\,{\tmsost} \nonumber \\
&&+{}\Biggl [{}\Biggl ({20\over 3}-{4}\,{\Ly}\Biggr ){}\,{\Lx^3}+{}\Biggl
({12}\,{\Ly^2}-{92\over 3}\,{\Ly}+{20\over 3}+{22\over
  3}\,{\LU}+{2}\,{\pi^2}\Biggr ){}\,{\Lx^2} \nonumber \\
&&+{}\Biggl ({}-{16}\,{\Ly^3}+{52}\,{\Ly^2}+{}\Biggl ({}-{44\over
  3}\,{\LU}-{80\over 3}-{16}\,{\pi^2}\Biggr ){}\,{\Ly}+{22\over
  3}\,{\LU}+{38\over 3}\,{\pi^2}\Biggr ){}\,{\Lx} \nonumber \\
&&+{8}\,{\Ly^4}-{104\over 3}\,{\Ly^3}+{}\Biggl ({16}\,{\pi^2}+{80\over
  3}+{44\over 3}\,{\LU}\Biggr ){}\,{\Ly^2}+{}\Biggl ({}-{104\over
  3}\,{\pi^2}-{44\over 3}\,{\LU}\Biggr ){}\,{\Ly} \nonumber \\
&&+{2\over
  3}\,{\pi^2}\,{}\Biggl ({3}\,{\pi^2}+{10}+{11}\,{\LU}\Biggr ){}\Biggr
]{}\,{\one}  \\
{X_{1;u}}&=&{}\Biggl [{\Lx^3}+{}\Biggl ({2}\,{\LU}-{4}\,{\Ly}+{1}\Biggr
){}\,{\Lx^2}+{}\Biggl ({6}\,{\Ly^2}+{}\Biggl ({}-{4}-{4}\,{\LU}\Biggr
){}\,{\Ly}+{\pi^2}+{2}\,{\LU}-{14\over 3}\Biggr ){}\,{\Lx} \nonumber \\
&&-{4}\,{\Ly^3}+{}\Biggl ({4}+{4}\,{\LU}\Biggr ){}\,{\Ly^2}+{}\Biggl
({}-{4}\,{\LU}+{28\over 3}-{4}\,{\pi^2}\Biggr
){}\,{\Ly}+{2}\,{\pi^2}\,{\LU}-{28\over 3}\,{\LU}+{\pi^2}\Biggr
]{}\,{\tpsost}\nonumber \\
&&+{}\Biggl [{1\over 3}\,{\Lx^3}+{}\Biggl ({}-{4\over 3}\,{\Ly}-{1}+{2\over
  3}\,{\LU}\Biggr ){}\,{\Lx^2}+{}\Biggl ({4\over 3}\,{\Ly^2}+{}\Biggl
({}-{4\over 3}\,{\LU}+{2}\Biggr ){}\,{\Ly}+{1\over
  3}\,{\pi^2}-{2}\,{\LU}\Biggr ){}\,{\Lx} \nonumber \\
&&+{1\over 3}\,{\pi^2}\,{}\Biggl ({2}\,{\LU}+{3}\Biggr ){}\Biggr
]{}\,{\tmsost} \nonumber \\
&&+{}\Biggl [{4\over 3}\,{\Lx^3}+{}\Biggl ({}-{16\over 3}\,{\Ly}+{8\over
  3}\,{\LU}+{4\over 3}\Biggr ){}\,{\Lx^2}+{}\Biggl ({8}\,{\Ly^2}+{}\Biggl
({}-{16\over 3}\,{\LU}-{16\over 3}\Biggr ){}\,{\Ly}+{4\over
  3}\,{\pi^2}+{8\over 3}\,{\LU}\Biggr ){}\,{\Lx} \nonumber \\
&&-{16\over 3}\,{\Ly^3}+{}\Biggl ({16\over 3}+{16\over 3}\,{\LU}\Biggr
){}\,{\Ly^2}+{}\Biggl ({}-{16\over 3}\,{\LU}-{16\over 3}\,{\pi^2}\Biggr
){}\,{\Ly}+{4\over 3}\,{\pi^2}\,{}\Biggl ({2}\,{\LU}+{1}\Biggr ){}\Biggr
]{}\,{\one}  \\
{X_{2;u}}&=&{}\Biggl [{}-{1\over 6}\,{\Lx^3}+{}\Biggl ({}-{1\over
  3}\,{\LU}+{4\over 3}\,{\Ly}-{10\over 9}\Biggr ){}\,{\Lx^2}+{}\Biggl
({}-{3}\,{\Ly^2}+{}\Biggl ({2}\,{\LU}+{40\over 9}\Biggr ){}\,{\Ly}-{7\over
  6}\,{\pi^2}-{31\over 9}\,{\LU}\Biggr ){}\,{\Lx} \nonumber \\
&&+{2}\,{\Ly^3}+{}\Biggl ({}-{2}\,{\LU}-{40\over 9}\Biggr
){}\,{\Ly^2}+{}\Biggl ({2}\,{\pi^2}+{62\over 9}\,{\LU}\Biggr
){}\,{\Ly}-{22\over 9}\,{\LU^2}-{\pi^2}\,{\LU}-{10\over 9}\,{\pi^2}\Biggr
]{}\,{\tpsost} \nonumber \\
&&+{}\Biggl [{}-{1\over 6}\,{\Lx^3}+{}\Biggl ({2\over 3}\,{\Ly}-{1\over
  3}\,{\LU}\Biggr ){}\,{\Lx^2}+{}\Biggl ({}-{2\over 3}\,{\Ly^2}-{1\over
  6}\,{\pi^2}+{2\over 3}\,{\Ly}\,{\LU}\Biggr ){}\,{\Lx}-{1\over
  3}\,{\pi^2}\,{\LU}\Biggr ]{}\,{\tmsost} \nonumber \\
&&+{}\Biggl [{}-{2\over 3}\,{\Lx^3}+{}\Biggl ({8\over 3}\,{\Ly}-{4\over
  3}\,{\LU}-{2\over 3}\Biggr ){}\,{\Lx^2}+{}\Biggl ({}-{4}\,{\Ly^2}+{}\Biggl
({8\over 3}\,{\LU}+{8\over 3}\Biggr ){}\,{\Ly}-{2\over 3}\,{\pi^2}-{4\over
  3}\,{\LU}\Biggr ){}\,{\Lx} \nonumber \\
&&+{8\over 3}\,{\Ly^3}+{}\Biggl ({}-{8\over 3}-{8\over 3}\,{\LU}\Biggr
){}\,{\Ly^2}+{}\Biggl ({8\over 3}\,{\LU}+{8\over 3}\,{\pi^2}\Biggr
){}\,{\Ly}-{2\over 3}\,{\pi^2}\,{}\Biggl ({2}\,{\LU}+{1}\Biggr ){}\Biggr
]{}\,{\one} \\
{X_{3;u}}&=&{}\Biggl [{1\over 18}\,{\Lx^2}+{}\Biggl ({}-{2\over 9}\,{\Ly}+{2\over 9}\,{\LU}\Biggr ){}\,{\Lx}-{4\over 9}\,{\Ly}\,{\LU}+{1\over 18}\,{\pi^2}+{2\over 9}\,{\LU^2}+{2\over 9}\,{\Ly^2}\Biggr ]{}\,{\tpsost}\\
{X_{4;u}}&=&{32\over 3}\,{\LU}\,{}\Biggl
[{\Lx}+{\Lx^2}-{2}\,{\Ly}-{2}\,{\Ly}\,{\Lx}+{2}\,{\Ly^2}+{\pi^2}\Biggr
]{}\,{\one} \nonumber \\
&&+{8\over 3}\,{\LU}\,{}\Biggl
[{3}\,{\pi^2}-{6}\,{\Ly}-{6}\,{\Ly}\,{\Lx}-{14}+{3}\,{\Lx^2}+{6}\,{\Ly^2}+{3}\,{\Lx}\Biggr
]{}\,{\tpsost} \nonumber \\
&&-{8\over 3}\,{\LU}\,{}\Biggl [{}-{\pi^2}+{2}\,{\Ly}\,{\Lx}-{\Lx^2}+{3}\,{\Lx}\Biggr ]{}\,{\tmsost}\\
{X_{5;u}}&=&{32\over 9}\,{\LU^2}\,{\tpsost}
\end{eqnarray}

\newpage
\section{One-loop master integrals}
\label{app:master_int}
In this appendix, we list the expansions for the one-loop box integrals in
$D=6-2\ep$.
We remain in the physical region $s>0$, $u,t < 0$, 
and write coefficients in terms of logarithms and polylogarithms that are
real in this domain.  More precisely, we use the notation of
Eqs.~(\ref{eq:xydef}) and~(\ref{eq:xydef1}) to define the arguments of the
logarithms and 
polylogarithms. The polylogarithms are defined as in
Eq.~(\ref{eq:lidef}).

We find that the box integrals have the expansion,
\begin{eqnarray}
\Bfin &=& \frac{ e^{\ep\gamma}
\Gamma  \left(  1+\epsilon \right)  \Gamma  
\left( 1-\epsilon \right) ^2 
 }{ 2s\Gamma  \left( 1-2 \epsilon  \right)   \left( 1-2 \epsilon  \right) } 
 \left(\frac{\mu^2}{s} \right)^{\ep}
  \Biggl\{
 \frac{1}{2}\lq\(\lnx-\lny\)^2+\pi^2 \rq\nonumber \\
&& 
 +2\ep \lq
 \Licx-\lnx\Libx-\frac{1}{3}\lnx^3-\frac{\pi^2}{2}\lnx \rq
\nonumber \\
&& 
-2\ep^2\Bigg[
\Lidx+\lny\Licx-\frac{1}{2}\lnx^2\Libx-\frac{1}{8}\lnx^4-\frac{1}{6}\lnx^3\lny+\frac{1}{4}\lnx^2\lny^2\nonumber
\\
&&\qquad
\qquad-\frac{\pi^2}{4}\lnx^2-\frac{\pi^2}{3}\lnx\lny-\frac{\pi^4}{45}\Bigg]
+ ( u \leftrightarrow t) \Biggr\} + \O{\ep^3},
\end{eqnarray}
and
\begin{eqnarray}
\label{eq:boxst}
{\rm Box}^6(s, t)&=&\frac{e^{\ep\gamma}\Gamma(1+\ep) \Gamma(1-\ep)^2}
{2 u\Gamma(1-2\ep)(1-2\ep)}\,\fu  \Biggl\{ \left(\Lx^2 +2 i\pi
\Lx\right)\nonumber \\
&&+\ep \Biggl[
\left(-2\Licx+2 \Lx \Libx  -\frac{2}{3} \Lx^3+2 \Ly \Lx^2-\pi^2 \Lx+2 \zeta_3
\right)\nonumber \\
&& \qquad \qquad +i\pi\left(2 \Libx +4 \Ly \Lx-\Lx^2-\frac{\pi^2}{3}\right)
\Biggr]\nonumber \\
&&+\ep^2 \Bigg[
\Biggl(2\Lidz+2\Lidy-2\Ly \Licx-2\Lx \Licy+(2\Lx\Ly-X^2-\pi^2)\Libx
\nonumber \\
&&\qquad+\frac{1}{3}\Lx^4-\frac{5}{3}\Lx^3\Ly+\frac{3}{2}\Lx^2\Ly^2+\frac{2}{3}\pi^2\Lx^2-2\pi^2\Lx\Ly+2\Ly\zeta_3+\frac{1}{6}\pi^4\Biggr)
\nonumber \\
&&\qquad \qquad + i\pi \biggl(
-2\Licx-2\Licy+2\Ly\Libx+\frac{1}{3}\Lx^3-2\Lx^2\Ly+3\Lx\Ly^2\nonumber \\
&& \qquad \qquad \qquad \qquad -\frac{\pi^2}{3}\Ly+2\zeta_3
\biggr)\Bigg] \Biggr\} + \O{\ep^3}.
\end{eqnarray}
${\rm Box}^6(s,u)$ is obtained from Eq.~(\ref{eq:boxst}) by exchanging $u$ and
$t$.

Finally, the one-loop bubble integral in $D=4-2 \epsilon$ dimensions 
is given by
\begin{equation} 
 \Bubl =\frac{ e^{\ep\gamma}\Gamma  \left(  1+\epsilon \right)  \Gamma  \left( 1-\epsilon \right) ^2 
 }{ \Gamma  \left( 2-2 \epsilon  \right)  \epsilon   } \fs.
\end{equation}


\newpage
\begin{thebibliography}{99}
\bibitem{BDK}Z. Bern, L. Dixon and D.A. Kosower, JHEP {\bf 0001} (2000) 027
[arXiv:hep-ph/0001001].
\bibitem{planarA}V.A. Smirnov, Phys. Lett. {\bf B460} (1999) 397
[arXiv:hep-ph/9905323].
\bibitem{nonplanarA}J.B. Tausk, Phys. Lett. {\bf B469} (1999) 225
[arXiv:hep-ph/9909506].
\bibitem{IBP}K.G.~Chetyrkin, A.L.~Kataev and F.V.~Tkachov, 
Nucl. Phys. {\bf B174} (1980) 345 ;\\
K.G.~Chetyrkin and F.V.~Tkachov, Nucl. Phys. {\bf B192} (1981) 159.
\bibitem{diffeq}T. Gehrmann and E. Remiddi, Nucl. Phys. {\bf B580} (2000)
485 [arXiv:hep-ph/9912329].
\bibitem{planarB}V.A. Smirnov and O.L. Veretin, Nucl. Phys. {\bf B566} (2000)
469 [arXiv:hep-ph/9907385].
\bibitem{nonplanarB}C. Anastasiou, T. Gehrmann, C. Oleari,
E. Remiddi and J.B. Tausk,  Nucl. Phys. {\bf B580} (2000) 577
[arXiv:hep-ph/0003261].
\bibitem{AGO3}C. Anastasiou, E.W.N. Glover and   C. Oleari,
 Nucl. Phys. {\bf B575} (2000) 416,  Erratum-ibid.\ {\bf B585} (2000) 763 
[arXiv:hep-ph/9912251].
\bibitem{onshell5}
T.~Gehrmann and E.~Remiddi, Nucl.\ Phys.\ {\bf B} (Proc.\ Suppl.)
{\bf 89} (2000) 251 [arXiv:hep-ph/0005232].
\bibitem{onshell6}
C.\ Anastasiou, J.B.\ Tausk and M.E.\ Tejeda-Yeomans, 
Nucl.\ Phys.\ {\bf B} (Proc.\ Suppl.) {\bf 89} (2000) 262
[arXiv:hep-ph/0005328].
\bibitem{BDG}Z. Bern, L. Dixon and A. Ghinculov, Phys. Rev. {\bf D63}
(2001) 053007 [arXiv:hep-ph/0010075].
\bibitem{qqQQ} C. Anastasiou, E.W.N. Glover, C. Oleari and M.E.
Tejeda-Yeomans, Nucl. Phys. {\bf B601} (2001) 318 [arXiv:hep-ph/0010212].
\bibitem{qqqq} C. Anastasiou, E.W.N. Glover, C. Oleari and M.E.
Tejeda-Yeomans, Nucl. Phys. {\bf B601} (2000) 341 [arXiv:hep-ph/0011094].
\bibitem{1loopsquare} C. Anastasiou, E.W.N. Glover, C. Oleari and M.E.
Tejeda-Yeomans, Phys. Lett. {\bf B506} (2001) 59 [arXiv:hep-ph/0012007].
\bibitem{qqgg} C. Anastasiou, E.W.N. Glover, C. Oleari and M.E.
Tejeda-Yeomans, Nucl. Phys. {\bf B605} (2001) 486 [arXiv:hep-ph/0101304].
\bibitem{gggg} E.W.N. Glover, C. Oleari and M.E.
Tejeda-Yeomans, Nucl. Phys. {\bf B605} (2001) 467 [arXiv:hep-ph/0102].
\bibitem{1loopgggg} E.W.N. Glover and M.E. Tejeda-Yeomans, JHEP {\bf 0105} 
(2001) 010 [arXiv:hep-ph/0104178].
\bibitem{ggpp} Z. Bern, A. De Freitas and L.J. Dixon, JHEP {\bf 0109} (2001)
037 [arXiv:hep-ph/0109078]. 
\bibitem{pppp} Z. Bern, A. De Freitas,  L.J. Dixon, A. Ghinculov and
H.L. Wong, JHEP {\bf 0111} (2001) 031 [arXiv:hep-ph/0109079]. 
\bibitem{gggg_BDD}
Z. Bern, A. De Freitas and L.J. Dixon, arXiv:hep-ph/0201161.
\bibitem{mi}
T.\ Gehrmann and E.\ Remiddi, Nucl.~Phys.~{\bf B601} (2001) 248
[arXiv:hep-ph/0008287];
{\bf B601} (2001) 287 [arXiv:hep-ph/0101124].
\bibitem{smirnov_offshell}
V.A.~Smirnov, Phys.\ Lett.\ {\bf B491} (2000) 130 [arXiv:hep-ph/007032]; 
{\bf B500} (2001) 330 [arXiv:hep-ph/0011056].
\bibitem{hpl}
E.\ Remiddi and J.A.M.\ Vermaseren, Int.\ J.\ Mod.\ Phys.\ {\bf A15}
(2000) 725 [arXiv:hep-ph/9905237].
\bibitem{zqqg} L.W. Garland, T. Gehrmann, E.W.N. Glover, A. Koukoutsakis and
  E. Remiddi,
arXiv:hep-ph/0112081.
\bibitem{smirnov_mass}V.A. Smirnov, Phys. Lett. {\bf B524} (2002) 129
[arXiv:hep-ph/0111160].
\bibitem{catani}S. Catani, Phys. Lett. {\bf B427} (1998) 161 [arXiv:hep-ph/9802439].
\bibitem{seymour} S. Catani and M.H. Seymour, Nucl. Phys. {\bf B485} (1997) 
291, erratum-ibid.{\bf B510} 503 [arXiv:hep-ph/9605323]. 
\bibitem{oneloopME} 
P. Aurenche, R. Baier, A. Douiri, M. Fontannaz and 
D. Schiff, 
Z. Phys. {\bf C24} (1984) 309; 
Nucl. Phys.  {\bf B286} (1987) 553. 
\bibitem{QGRAF}P. Nogueira, J.Comput.Phys. {\bf 105} (1993) 279. 
\bibitem{AGO2}C. Anastasiou, E.W.N. Glover and   C. Oleari,
 Nucl. Phys. {\bf B572} (2000) 307 [arXiv:hep-ph/9907494].
\bibitem{xtri}R.J. Gonsalves, Phys. Rev. {\bf D28} (1983) 1542;\\
G. Kramer and B. Lampe, J. Math. Phys. {\bf 28} (1987) 945.
\bibitem{tc}J.M. Campbell and E.W.N. Glover, Nucl. Phys. {\bf B527} (1998)
264 [arXiv:hep-ph/9710255];\\
S. Catani and M. Grazzini, Phys. Lett. {\bf 446B} (1999) 143 
[arXiv:hep-ph/9810389];\\
S. Catani and M. Grazzini, Nucl. Phys. {\bf B570} (2000) 287 [arXiv:hep-ph/9908523].
\bibitem{ds}F.A. Berends and W.T. Giele, Nucl. Phys. {\bf B313} (1989) 595;\\
S. Catani, in Proceedings of the workshop on {\em New Techniques for Calculating
Higher Order QCD Corrections}, report ETH-TH/93-01, Zurich (1992).
\bibitem{sone}Z. Bern, V. Del Duca and C.R. Schmidt, Phys. Lett. {\bf 445B}
(1998) 168 [arXiv:hep-ph/9810409];\\
Z. Bern, V. Del Duca, W.B. Kilgore and C.R. Schmidt, Phys. Rev. {\bf D60} (1999)
116001 [arXiv:hep-ph/9903516];\\
S. Catani and M. Grazzini, Nucl. Phys. {\bf B591} 435 [arXiv:hep-ph/0007142].
\bibitem{cone}Z. Bern, L. Dixon, D.C. Dunbar and D.A. Kosower,
Nucl. Phys. {\bf B425} (1994) 217 [arXiv:hep-ph/9403226];\\
D.A. Kosower, Nucl. Phys. {\bf B552} (1999) 319;\\
D.A. Kosower and P. Uwer, Nucl. Phys. {\bf B563} (1999) 477 [arXiv:hep-ph/9903515].
\bibitem{aude} A. Gehrmann-De Ridder and E.W.N. Glover, Nucl. Phys. {\bf B517}
(1998) 269 [arXiv:hep-ph/9707224].
\bibitem{grazzini}S. Catani, D. de Florian and M. Grazzini,
JHEP {\bf 0105} (2001) 025 [arXiv:hep-ph/0102227]; 
R.V. Harlander and W.B. Kilgore, Phys. Rev. {\bf D 64} (2001) 013015 
[arXiv:hep-ph/0102241] ;
R.V. Harlander, Phys. Lett. {\bf B 492} (2000) 74 [arXiv:hep-ph/0007289].
\bibitem{kolbig}
K.S.~K\"olbig, J.A.~Mignaco and E.~Remiddi,
B.I.T.\ {\bf 10} (1970) 38.
\bibitem{moms1}  S.A. Larin, T. van Ritbergen, and J.A.M. Vermaseren,
                 Nucl.\ Phys.\ {\bf B427} (1994) 41;\\
                 S.A. Larin, P. Nogueira, T. van Ritbergen, and J.A.M.
                 Vermaseren, Nucl.\ Phys.\ {\bf B492} (1997) 338
                 [arXiv:hep-ph/9605317].
\bibitem{moms2}  A. Retey and J.A.M. Vermaseren, Nucl. Phys. {\bf B604}
(2001) 281 [arXiv:hep-ph/0007294].
\bibitem{Gra1}   J. A. Gracey, Phys.\ Lett.\ {\bf B322} (1994) 141
[arXiv:hep-ph/9401214].
\bibitem{NV}    W.L. van Neerven and A. Vogt, Nucl.\ Phys.\ {\bf B568}
                 (2000) 263 [arXiv:hep-ph/9907472]; Nucl. Phys. {\bf B588}
		 (2000) 345 [arXiv:hep-ph/0006154]. 
\bibitem{NVplb}  W.L. van Neerven and A. Vogt, Phys.\ Lett.\ {\bf B490}
                 (2000) 111 [arXiv:hep-ph/0007362].

\bibitem{MRS} A.D. Martin, R.G. Roberts, W.J. Stirling and R.S. Thorne,
Eur.\ Phys.\ J.\ {\bf C18} (2000) 117 [arXiv:hep-ph/0007099].

\end{thebibliography}
\end{document}